\documentclass[hyper,letterpaper]{JHEP3}

\usepackage{epsfig}
\usepackage{amsbsy}
\usepackage{varioref}
\usepackage{pifont}

\newcommand{\dfrac}[2]{\frac{\strut \displaystyle{#1}}{\strut \displaystyle{#2}}}

\title{Electroweak Corrections and Unitarity\\
in Linear Moose Models}

\author{R. Sekhar Chivukula and Elizabeth H. Simmons\\
Department of Physics and Astronomy, Michigan State University\\
East Lansing, MI 48824, USA\\
	E-mail: \email{sekhar@msu.edu, esimmons@msu.edu}}
	
\author{Hong-Jian He\\
Department of Physics, University of Texas\\
Austin, TX 78712, USA\\
	E-mail: \email{hjhe@physics.utexas.edu}}

\author{Masafumi Kurachi\\
Department of Physics, Nagoya University\\
Nagoya 464-8602, Japan\\
	E-mail:\email{kurachi@eken.phys.nagoya-u.ac.jp}}

\author{Masaharu Tanabashi\\
Department of Physics, Tohoku University\\
Sendai 980-8578, Japan\\
	E-mail:\email{tanabash@tuhep.phys.tohoku.ac.jp}}

\abstract{
We calculate the form of the corrections to 
the electroweak interactions in the class of Higgsless models which can be 
``deconstructed'' to a  chain of $SU(2)$ gauge groups adjacent to a chain of $U(1)$ gauge groups, 
and with the fermions coupled to any single $SU(2)$ group and to any single 
$U(1)$ group along the chain. The primary advantage of  our technique is 
that the size of corrections to electroweak processes
can be directly related to the spectrum of vector bosons (``KK modes"). In Higgsless models, this spectrum 
is constrained by unitarity.  Our methods also allow for arbitrary background 5-D geometry,
spatially dependent gauge-couplings, and brane kinetic energy terms. 
We find that, due to the size of corrections to electroweak
processes in any {\it unitary} theory, Higgsless models with localized fermions are disfavored by
precision electroweak data. 
Although we stress our results as they apply to continuum Higgsless
5-D models, they apply to any linear moose model including those
with only a few extra vector bosons.
Our calculations of electroweak corrections 
also apply directly to the electroweak gauge sector of 5-D theories with a bulk
scalar Higgs boson; the constraints arising from unitarity do not apply in this case.
}


\preprint{MSUHEP-041011\\
TU-733\\
DPNU-04-18}

\begin{document}


\section{Introduction}

The mechanism that spontaneously breaks the electroweak gauge symmetry remains unknown.  One novel solution to the puzzle is embodied in the ``Higgsless'' models \cite{Csaki:2003dt}, which are based on five-dimensional gauge theories compactified on an interval.  These models 
achieve unitarity of electroweak boson self-interactions 
through the exchange of a tower of massive vector bosons 
\cite{SekharChivukula:2001hz,Chivukula:2002ej,Chivukula:2003kq}, 
rather than the exchange of a scalar Higgs boson \cite{Higgs:1964ia}.

In this paper, using deconstruction \cite{Arkani-Hamed:2001ca,Hill:2000mu}, we calculate the form of the corrections to the electroweak interactions in a large class of these models in which the fermions are localized in the extra dimension. Our analysis applies to any Higgsless model which can be deconstructed to a  chain of $SU(2)$ gauge groups adjacent to a chain of $U(1)$ gauge groups,  with the fermions coupled to any single $SU(2)$ group and to any single 
 $U(1)$ group along the chain.\footnote{Recently, it has been proposed that the size of
corrections to electroweak processes may be reduced by allowing for delocalized
fermions  \protect{\cite{Cacciapaglia:2004rb,Foadi:2004ps}}. Here we restrict our attention
to the case of localized fermions.}  The analyses presented here extend and
generalize those we have presented previously \cite{Chivukula:2004pk,Chivukula:2004af}.

The primary advantage of  our technique is that the size of corrections to electroweak processes
can be directly related to the spectrum of vector bosons (``KK modes") which, in Higgsless models,
is constrained by unitarity. In addition, our results allow for arbitrary background 5-D geometry,
spatially dependent gauge-couplings, and brane kinetic energy terms. 
We find that, due to the size of corrections to electroweak
processes in any unitary theory, Higgsless models with localized fermions are disfavored by
precision electroweak data. 

We find that the case (which we will designate ``Case I") in which the fermions' hypercharge interactions are with the $U(1)$ group at the interface between the $SU(2)$ and $U(1)$ groups are of particular phenomenological interest.   Previous results \cite{Chivukula:2004pk} for the sub-set of Case I models in which fermions couple only to the $SU(2)$ group at the left-most end of the chain are recovered as a special limit of our general expressions, and the results of our analysis for Case I was quoted
in \cite{Chivukula:2004af}. We also examine the subset of Case I models where fermions couple only to the $SU(2)$  group  at the interface; this is an extension of the Generalized BESS model \cite{Chivukula:2003wj,Casalbuoni:2004id}.  We find that in this limit, $S$ and $T$ take on their minimum values  and the quantity  $S - 4 \cos^2\theta_W T$ vanishes to leading order in the ratio $M_Z^2/{\cal M}_{Zz}^2$ where ${\cal M}_{Zz}$ is the mass of the extra $Z'$ bosons in the model. 

Although we stress our results as they apply to continuum Higgsless
5-D models, they apply also far from the continuum limit
when only a few extra vector bosons are present. As such, these
results form a generalization of phenomenological analyses \cite{Chivukula:2003wj} of models of extended  electroweak gauge symmetries \cite{Casalbuoni:1985kq,Casalbuoni:1996qt,Casalbuoni:2004id}  motivated by models of hidden local symmetry \cite{Bando:1985ej,Bando:1985rf,Bando:1988ym,Bando:1988br,Harada:2003jx}.
Our calculations also apply directly to the electroweak gauge sector of 5-D theories with a bulk
scalar Higgs boson, although the constraints arising from unitarity no longer apply.

The results presented here are complementary to, and more general than,
 the analyses of the phenomenology of these modes in the continuum
\cite{Csaki:2003zu,Nomura:2003du,Barbieri:2003pr,Davoudiasl:2003me,Foadi:2003xa,Burdman:2003ya,Davoudiasl:2004pw,Barbieri:2004qk,Hewett:2004dv}. Recently, 
using deconstruction,  Perelstein \cite{Perelstein:2004sc} has argued that the 
higher-order corrections expected
to be present in any QCD-like ``high-energy'' completion of a Higgsless theory are
also likely to be large. In this work, we compute the tree-level corrections expected
independent of the form of the high-energy completion.

In the next three sections, we introduce the models we will analyze, 
set notation for the correlation functions and vector-boson mass matrices, and
carefully specify the electroweak parameters we will compute. In section 5, we show
that one correlation function may be computed quite directly in the most general
model. In sections 6
and 7, we discuss the correlation functions and phenomenology of ``Case I'' in detail, and then generalize this analysis
in sections 8 and 9 for a moose with fermions coupled to an arbitrary $U(1)$ group. The primary results of this paper, the form of
electroweak corrections in the most general moose model with localized
fermions, are summarized in section 9. In section 10, we demonstrate that the
size of the electroweak corrections is related to the unitarity of the theory, and show
that these models are disfavored by precision electroweak data. Following a short
discussion and summary, the appendices present various technical and notational
elaborations of the discussion given in the main body of the paper. 

\section{The Model and Its Relatives}

\EPSFIGURE[t]{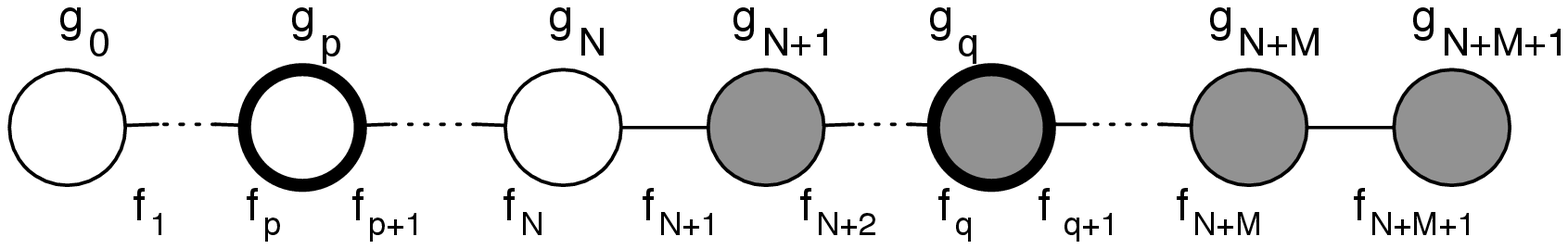,width=0.9\textwidth}
{Moose diagram for the class of models analyzed in this paper. $SU(2)$ gauge groups are shown as open circles; $U(1)$ gauge groups as shaded circles. The fermions couple to gauge
groups $p$ $(0 \leq p \leq N$) and $q$ ($N+1 \leq q \leq N+M+1$).  The values of the gauge couplings $g_i$ and $f$-constants $f_i$ are arbitrary.
\label{fig:TheMoose}}

The model we analyze, shown diagrammatically (using ``moose notation'' \cite{Georgi:1986hf,Arkani-Hamed:2001ca}) in Fig. \ref{fig:TheMoose}, incorporates an
$SU(2)^{N+1} \times U(1)^{M+1}$ gauge group, and $N+1$ 
nonlinear $(SU(2)\times SU(2))/SU(2)$ sigma models adjacent to $M$
$(U(1) \times U(1))/U(1)$ sigma models in which the global symmetry groups 
in adjacent sigma models are identified with the corresponding factors of the gauge group.
The Lagrangian for this model at $O(p^2)$ is given by
\begin{equation}
  {\cal L}_2 =
  \frac{1}{4} \sum_{j=1}^{N+M+1} f_j^2 \mbox{tr}\left(
    (D_\mu U_j)^\dagger (D^\mu U_j) \right)
  - \sum_{j=0}^{N+M+1} \dfrac{1}{2g_j^2} \mbox{tr}\left(
    F^j_{\mu\nu} F^{j\mu\nu}
    \right),
\label{lagrangian}
\end{equation}
with
\begin{equation}
  D_\mu U_j = \partial_\mu U_j - i A^{j-1}_\mu U_j 
                               + i U_j A^{j}_\mu,
\end{equation}
where all  gauge fields $A^j_\mu$ $(j=0,1,2,\cdots, N+M+1)$ are dynamical. The first
$N+1$ gauge fields ($j=0,1,\ldots, N$) correspond to $SU(2)$ gauge groups; the other $M+1$ gauge
fields ($j=N+1, N+2, \ldots, N+M+1$) correspond to $U(1)$ gauge groups.  The symmetry breaking between
the $A^{N}_\mu$ and $A^{N+1}_\mu$ follows an $SU(2)_L \times SU(2)_R/SU(2)_V$ symmetry
breaking pattern with the $U(1)$ embedded as the $T_3$-generator of $SU(2)_R$. In what follows,
we will denote $N+M$ by $K$ for brevity.

The fermions in this model take their weak interactions from the $SU(2)$ group at $j=p$ and their hypercharge interactions from the $U(1)$ group with $j=q$.  The neutral current couplings to the fermions are thus written as
\begin{equation}
J^\mu_3 A^p_\mu + J^\mu_Y A^{q}_\mu~,
\label{eq:current}
\end{equation}
while the charged current couplings arise from
\begin{equation}
{1 \over \sqrt{2}} J^\mu_{\pm} A^{p\mp}_\mu~.
\end{equation}
We allow $p$ to assume any value $0 \leq p \leq N$; we consider the special "Case I" mooses in which $q = N+1$ and also the more general situation in which $q$ can take on any value in the range $N+1 \leq q \leq K+1$.

Our analysis proceeds for arbitrary values of the gauge couplings and $f$-constants, 
and therefore allow for arbitrary background 5-D geometry,
spatially dependent gauge-couplings, and brane kinetic energy terms for the gauge-bosons. 

\section{Notation: Propagators, Mass-Squared Matrices,\\
 and Correlation Functions}

All four-fermion processes, including those relevant for the electroweak phenomenology of our model, depend on the neutral gauge field
propagator matrix
\begin{equation}
D^Z(Q^2) \equiv \left[ Q^2\, {\cal I} + M^2_{Z}\right]^{-1}~,
\end{equation}
and the charged propagator matrix
\begin{equation}
{D}^W(Q^2) \equiv \left[ Q^2\, {\cal I} + {M}^2_{W}\right]^{-1}~.
\end{equation}
Here, $M_{Z}^2$ and ${M}_W^2$ are, respectively, the mass-squared matrices for the neutral and charged gauge bosons and ${\cal I}$ is the identity matrix.  Consistent with \cite{Chivukula:2004pk}, $Q^2 \equiv -q^2$ refers to the
euclidean momentum. 

Recalling that fermions are charged under only a single SU(2) gauge group (at $j=p$ where $0 \leq p \leq N$) and a single U(1) group (at $j=q$ where $N+1 \leq q \leq K+1$), neutral current four-fermion processes may be derived from the Lagrangian
\begin{equation}
{\cal L}_{nc} = - {1\over 2} g^2_p\, D^Z_{p,p}(Q^2) J^\mu_3 J_{3\mu}
- g_p g_{q}\, D^Z_{p,q}(Q^2) J^\mu_3 J_{Y\mu}
-  {1\over 2} g^2_{q}\, D^Z_{q,q}(Q^2) J^\mu_Y J_{Y\mu}~,
\label{nclagrangian}
\end{equation}
and charged-current process from
\begin{equation}
{\cal L}_{cc} = - {1\over 2} g^2_p\, {D}^W_{p,p}(Q^2) J^\mu_+ J_{-\mu}~.
\label{cclagrangian}
\end{equation}
where $D_{i,j}$ is the (i,j) element of the appropriate gauge field propagator matrix.

The neutral vector meson mass-squared matrix is of dimension $(K+2) \times (K+2)$ 
\begin{equation}
{\tiny
M_{Z}^2 = {1\over 4}
\left(
\begin{array}{c|c|c|c|c|c}
g^2_0 f^2_1& -g_0 g_1 f^2_1 & & &  \\ \hline
-g_0 g_1 f^2_1  & g^2_1(f^2_1+f^2_2) & -g_1 g_2 f^2_2 & &   \\ \hline
 & \ddots & \ddots & \ddots &  \\ \hline
 & & -g^{}_{N} g^{}_{N+1} f^2_{N+1} & g^2_{N+1}(f^2_{N+1} + f^2_{N+2}) & -g^{}_{N+1} g^{}_{N+2} f^2_{N+2} &   \\ \hline
  & & & -g^{}_{N+1} g^{}_{N+2} f^2_{N+2} & g^2_{N+2}(f^2_{N+2}+f^2_{N+3}) & \ddots  \\ \hline
 & & & & \ddots & \ddots \\
\end{array}
\right).
}
\label{eq:neutralmatrix}
\end{equation}
and the charged current vector bosons' mass-squared matrix is the upper $(N+1)  \times (N+1) $ dimensional block of the neutral current $M_{Z}^2$ matrix
\begin{equation}
{\tiny
{M}_{W}^2 = {1\over 4}
\left(
\begin{array}{c|c|c|c|c|c}
g^2_0 f^2_1& -g_0 g_1 f^2_1 & & &  \\ \hline
-g_0 g_1 f^2_1  & g^2_1(f^2_1+f^2_2) & -g_1 g_2 f^2_2 & &   \\ \hline
 & -g_1 g_2 f_2^2 & g^2_2(f^2_2+f^2_3) & -g_2 g_3 f^2_3 &  \\ \hline
 & & \ddots & \ddots & \ddots &   \\ \hline
  & & & -g^{}_{N-2} g^{}_{N-1} f^2_{N-1} & g^2_{N-1}(f^2_{N-1}+f^2_{N}) & -g^{}_{N-1} g^{}_{N} f^2_{N} \\ \hline
 & & & & -g^{}_{N-1} g^{}_{N} f^2_{N} & g^2_{N}(f^2_{N}+f^2_{N+1})\\
\end{array}
\right).
}
\label{chargedmatrix}
\end{equation}
The neutral vector boson mass matrix (\ref{eq:neutralmatrix}) 
is of a familiar form that has a vanishing determinant, due to a zero eigenvalue.
Physically, this corresponds to a massless neutral gauge field -- the photon.
The non-zero eigenvalues of $M^2_{Z}$
are labeled by ${\mathsf m}^2_{Zz}$ ($z=0,1,2,\cdots, K$), while
those of ${M}^2_W$ are labeled by ${\mathsf m}^2_{Ww}$ ($w=0, 1,2,\cdots, N$). 
The lowest massive eigenstates corresponding to eigenvalues ${\mathsf m}^2_{Z0}$ and 
${\mathsf m}^2_{W0}$ are, respectively, identified as the usual $Z$ and $W$ bosons.
We will generally refer to  these last eigenvalues by their conventional symbols $M^2_Z,\, M^2_{W}$; the distinction between these and the corresponding mass matrices should be clear from
context.

It is useful to define an  $M\times M$ matrix  
${\cal M}_M$ which encompasses all of the $U(1)$ groups corresponding to $j > N+1$ (i.e., all except the left-most $U(1)$ group in the linear moose).
\begin{eqnarray}
  \lefteqn{
      {\cal M}_M^2 = 
  } \nonumber\\
 & & {\tiny
{1\over 4}
\left(
\begin{array}{c|c|c|c|c|c}
g_{N+2}^2 (f_{N+2}^2+f_{N+3}^2)   & -g^{}_{N+2} g^{}_{N+3} f_{N+3}^2      &
   &  &  & \\ \hline
-g^{}_{N+2} g^{}_{N+3} f_{N+3}^2        & g_{N+3}^2 (f_{N+3}^2+f_{N+4}^2) & 
-g^{}_{N+3} g^{}_{N+4} f_{N+4}^2  &  &  & \\ \hline
                                  & -g^{}_{N+3} g^{}_{N+4} f_{N+4}^2      &
 g_{N+4}^2 (f_{N+4}^2+f_{N+5}^2)  & -g^{}_{N+4} g^{}_{N+5} f_{N+5}^2      &
                                  & \\ \hline
 & & \ddots & \ddots & \ddots  &   \\ \hline
 & & & -g^{}_{K-1} g^{}_{K} f_K^2         &
          g_K^2(f_K^2+f_{K+1}^2) &  -g^{}_K g^{}_{K+1} f_{K+1}^2       \\ \hline
 & & & & -g^{}_K g^{}_{K+1} f_{K+1}^2  &   g_{K+1}^2 f^2_{K+1}         \\
\end{array}
\right)~,
}
\nonumber\\
 & &
\label{eq:mass_matrixM}
\end{eqnarray}
(where $K \equiv N+M$).
 The neutral-current matrix (\ref{eq:neutralmatrix}) can be written more compactly as 
\begin{equation}
{\tiny
  M_{Z}^2 = \left(
    \begin{array}{cc|c|cc}
      \multicolumn{2}{c|}{{ M_{W}^2}} & & 
      \multicolumn{2}{c}{} 
      \\ 
      \multicolumn{2}{c|}{} & - g^{}_{N} g^{}_{N+1} f_{N+1}^2/4 & 
      \multicolumn{2}{c}{} 
      \\
      \hline
      \phantom{wwwwww}
      & - g^{}_{N} g^{}_{N+1} f_{N+1}^2/4 
      &   g_{N+1}^2 (f_{N+1}^2 + f_{N+2}^2)/4
      & - g^{}_{N+1} g^{}_{N+2} f_{N+2}^2/4
      & 
      \phantom{wwwwww}\\
      \hline
      \multicolumn{2}{c|}{} & - g^{}_{N+1} g^{}_{N+2} f_{N+2}^2/4 & &
      \\
      \multicolumn{2}{c|}{} & & 
      \multicolumn{2}{c}{{\cal M}_{M}^2}
    \end{array}
  \right).
}
\label{eq:mass_matrixN2}
\end{equation}
The eigenvalues of ${\cal M}^2_M$,
are labeled by ${\mathsf m}^2_{m}$ ($m=1,2,\cdots, M$), and the 
associated propagator is denoted as  
$ {\cal D}^M(Q^2) \equiv \left[ Q^2 {\cal I} + {\cal M}^2_M\right]^{-1}$.

Generalizing the usual mathematical
notation for ``open'' and ``closed'' intervals, we may denote the
neutral-boson mass matrix $M^2_Z$ as $M^2_{[0,K+1]}$ --- {\it i.e.}
it is the mass matrix for the entire moose running from site $0$ to site $K+1$ including
the gauge couplings of both endpoint groups. Analogously, the charged-boson mass matrix $M^2_W$ is
$M^2_{[0,N+1)}$ --- it is the mass matrix for the moose running from site $0$ to link
$N+1$, but not including the gauge couping at site $N+1$.
This notation will be useful in thinking about the properties of the various sub-matrices 
and in keeping track of the  end values of sums over eigenvalues.
Using this notation, we can write ${\cal M}^2_M$ as
\begin{equation}
{\cal M}^2_M = M^2_{(N+1,K+1]}~.
\end{equation}
 We can also define other useful submatrices  ${\cal M}_i^2$ and their eigenvalues ${\mathsf m}_i^2$ as
\begin{eqnarray}
{\cal M}^2_p & = & M^2_{[0,p)} \qquad {\mathsf m}^2_{\hat{p}}\ ({\hat{p}}=1,2,\cdots, p)\\
{\cal  M}^2_r & = & M^2_{(p,N+1)} \qquad {\mathsf m}_r^2\  (r=p+1,p+2,\cdots, N)\\
{\cal  M}^2_q & = & M^2_{(q,K+1]} \qquad {\mathsf m}^2_{\hat{q}}\ (\hat{q}=q+1,\cdots, K+1) \\
 {\cal M}^2_{L} &=& M^2_{(p,K+1]} \qquad {\mathsf m}^2_l\ (l=0,1,\cdots, K-p) .
\end{eqnarray}
The propagator matrix related to each ${\cal M}^2_i$ is written $ {\cal D}^i(Q^2) \equiv \left[ Q^2 {\cal I} + {\cal M}^2_i\right]^{-1}$. Note that ${\cal M}^2_M$ is the same as ${\cal M}^2_{q = N+1}$.   We will show that the lowest  ($l = 0$) eigenvalue of the matrix ${\cal M}^2_L$ is typically light (comparable to $M^2_{W,Z}$), and we will denote it $M^2_L$.  For completeness, the matrices ${\cal M}^2_{p,q,r, L}$ are written out  explicitly in Appendix A.

The propagator matrix elements related to $M_Z^2$ and $M_W^2$ may be written in a spectral decomposition in terms of the mass eigenstates as follows:
\begin{eqnarray}
  g^2_p D^Z_{p,p}(Q^2)\equiv [G_{\rm NC}(Q^2)]_{WW} &=& 
    \dfrac{[\xi_\gamma]_{WW}}{Q^2}
   +\dfrac{[\xi_Z]_{WW}}{Q^2 + M_Z^2}
   +\sum_{z=1}^{K} \dfrac{[\xi_{Zz}]_{WW}}
                                  {Q^2 + {\mathsf m}_{Zz}^2},
\label{eq:NC_WW}  
  \\
  g_p g_{q} D^Z_{p,q}(Q^2) \equiv [G_{\rm NC}(Q^2)]_{WY} &=& 
    \dfrac{[\xi_\gamma]_{WY}}{Q^2}
   +\dfrac{[\xi_Z]_{WY}}{Q^2 + M_Z^2}
   +\sum_{z=1}^{K} \dfrac{[\xi_{Zz}]_{WY}}
                                  {Q^2 + {\mathsf m}_{Zz}^2},
\label{eq:NC_WY}  
  \\
  g^2_{q} D^Z_{q,q}(Q^2)\equiv [G_{\rm NC}(Q^2)]_{YY} &=& 
    \dfrac{[\xi_\gamma]_{YY}}{Q^2}
   +\dfrac{[\xi_Z]_{YY}}{Q^2 + M_Z^2}
   +\sum_{z=1}^{K} \dfrac{[\xi_{Zz}]_{YY}}
                                  {Q^2 + {\mathsf m}_{Zz}^2},
\label{eq:NC_YY}  
  \\
  g^2_p {D}^W_{p,p}(Q^2)\equiv [G_{\rm CC}(Q^2)]_{WW} &=& 
   \dfrac{[\xi_W]_{WW}}{Q^2 + M_W^2}
   +\sum_{w=1}^{N} \dfrac{[\xi_{Ww}]_{WW}}
                                  {Q^2 + {\mathsf m}_{Ww}^2},
\label{eq:CC_WW}  
\end{eqnarray}
All poles should be simple (i.e. there should be no degenerate mass eigenvalues) because, in the continuum limit, we are analyzing a self-adjoint operator on a finite interval.

The pole residues $\xi$ directly show the contributions of the various weak bosons to four-fermion processes.  We can also write the couplings of the $Z$ and $W$ bosons in terms of the residues
\begin{equation}
 ( \sqrt{[\xi_Z]_{WW}} \, J_{3\mu} - \sqrt{[\xi_Z]_{YY}}\,  J_{Y\mu} )\, Z^\mu
 \label{eq:zzcoup}
\end{equation}
\begin{equation}
 \sqrt{ \dfrac{[\xi_W]_{WW}}{2}} \, J^\mp_\mu \,W^{\pm\mu}
\end{equation}
and see how the existence of the heavy gauge bosons alters these couplings from their standard model values.  

In each case, the $W$ and $Z$ couplings will approach their tree-level standard model values  in the limit ${\mathsf m}^2_{Z,{z>0}}, {\mathsf m}^2_{W,{w>0}} \to  \infty$.
From a phenomenological point of view, therefore, it is particularly interesting to consider scenarios in which the only light gauge bosons are the photon, $W$, and $Z$, i.e. with all of the extra charged and neutral gauge bosons being much heavier.  Ratios of physical masses such as $M_{Z}^2 / {\mathsf m}_{Z{z}}^2$ will, then, be small.  Phenomenological constraints will also require that 
the mass-squared eigenvalues ${\mathsf m}_{m}^2$, ${\mathsf m}_{\hat{p}}^2$, ${\mathsf m}_{{r}}^2$, ${\mathsf m}_{\hat{q}}^2$, and ${\mathsf m}^2_{l>0}$ (which do not directly correspond
to the masses of any physical particles) be large.   
We will  find it useful to define the following sums over the heavy masses\footnote{Note that the sums in $\Sigma_Z$, $\Sigma_W$ and $\Sigma_L$ start, respectively, from $z=1$, $w=1$, and $l = 1$, excluding the lightest non-zero mass eigenvalue in each case.} for these phenomenological discussions:
\begin{equation}
\Sigma_{Z} \equiv \sum_{z=1}^K {1\over {\mathsf m}^2_{Zz}}\ ,\ \ \ \ \ 
\Sigma_{W} \equiv  \sum_{w=1}^N {1\over {\mathsf m}^2_{Ww}}\ ,\ \ \ \ \ 
\Sigma_{M}  \equiv \sum_{m=1}^M {1\over {\mathsf m}^2_{m}}\ ,
\label{eq:sig-def}
\end{equation}
\begin{equation}
\Sigma_{p} \equiv \sum_{\hat{p}=1}^{p} {1\over {\mathsf m}^2_{\hat{p}}}\ ,\ \ \ \ \ 
\Sigma_{q} \equiv  \sum_{\hat{q}=q+1}^{K+1} {1\over {\mathsf m}^2_{\hat{q}}}\ , \ \ \ \ 
\Sigma_{r} \equiv \sum_{r=p+1}^{N} {1\over {\mathsf m}^2_r}\, \ \ \ \ 
\Sigma_{L} \equiv \sum_{l=1}^{K-p} {1\over {\mathsf m}^2_l}~.
\label{eq:sig-def-a}
\end{equation}
Note that for $p=0$ we have $\Sigma_p = 0$ and likewise for $q=K+1$ we have $\Sigma_q = 0$; this makes sense since the fermions then couple to the groups at the ends of the moose and the respective  matrix ${\cal M}_p$ or ${\cal M}_q$ does not exist.  When $q=N+1$ we have 
$\Sigma_q = \Sigma_{ M}$.  Thus we expect that setting $\Sigma_p = 0$, $\Sigma_q = 
\Sigma_{M}$ the results of our general analysis will recover the results in reference \cite{Chivukula:2004pk}.

Since the neutral bosons couple to only two currents, $J^\mu_3$ and $J^\mu_Y$, 
the three sets of residues in equations (\ref{eq:NC_WW})--(\ref{eq:NC_YY}) must be related. 
Specifically, they satisfy the $K+1$ consistency conditions,
\begin{equation}
  [\xi_Z]_{WW} [\xi_Z]_{YY}
  = \left([\xi_Z]_{WY}\right)^2, \qquad
  [\xi_{Z{z}}]_{WW} [\xi_{Z{z}}]_{YY}
  = \left([\xi_{Z{z}}]_{WY}\right)^2 .
\label{consistency}
\end{equation}
In the case of the photon, charge universality further implies
\begin{equation}
  e^2 = [\xi_\gamma]_{WW} = [\xi_\gamma]_{WY} = [\xi_\gamma]_{YY}.
\label{eq:universality}
\end{equation}


 \section{Electroweak Parameters}

Our goal is to analyze four-fermion electroweak processes in the general
linear moose model. As we have shown in \cite{Chivukula:2004af}, the most
 general amplitude for low-energy four-fermion neutral weak current processes in
any ``universal'' model \cite{Barbieri:2004qk} may be written 
as\footnote{See \cite{Chivukula:2004af} for a discussion of the correspondence 
between the ``on-shell'' parameters defined here, and the zero-momentum
parameters defined in  \protect{\cite{Barbieri:2004qk}}.  Note that $U$ is shown in \cite{Chivukula:2004af} to be zero to the order we consider in this paper.}
\begin{eqnarray}
-{\cal A}_{NC} = e^2 {{\cal Q}{\cal Q}' \over Q^2} 
& + &
\dfrac{(I_3-s^2 {\cal Q}) (I'_3 - s^2 {\cal Q}')}
	{\left({s^2c^2 \over e^2}-{S\over 16\pi}\right)Q^2 +
		{1\over 4 \sqrt{2} G_F}\left(1+{\alpha \delta \over 4 s^2 c^2}-\alpha T\right)
		} 
\label{eq:NC-M} \\ \nonumber & \ \ & \\
&+&
\sqrt{2} G_F \,{\alpha \delta\over  s^2 c^2}\, I_3 I'_3 
+ 4 \sqrt{2} G_F  \left( \Delta \rho - \alpha T\right)({\cal Q}-I_3)({\cal Q}'-I_3')~,
\nonumber 
\end{eqnarray}
and the matrix element for charged currents by 
\begin{eqnarray}
  - {\cal A}_{\rm CC}
  =  \dfrac{(I_{+} I'_{-} + I_{-} I'_{+})/2}
             {\left(\dfrac{s^2}{e^2}-\dfrac{S}{16\pi}\right)Q^2
             +{1\over 4 \sqrt{2} G_F}\left(1+{\alpha \delta \over 4 s^2 c^2}\right)
            }
        + \sqrt{2} G_F\, {\alpha  \delta\over s^2 c^2} \, {(I_{+} I'_{-} + I_{-} I'_{+}) \over 2}~.
\label{eq:CC-M}
\end{eqnarray}
Here $I^{(\prime)}_a$ and ${\cal Q}^{(\prime)}$ are weak isospin and charge
of the corresponding fermion, $\alpha = e^2/4\pi$, $G_F$ is the usual Fermi constant,
and the weak mixing angle, as defined by the on-shell $Z$ coupling, is denoted by $s^2$ 
($c^2\equiv 1-s^2$).

We can read off the forms of the Z-pole and W-pole residues in the four-fermion amplitudes (\ref{eq:NC-M}) and (\ref{eq:CC-M}) in terms of the parameters $S$, $T$, and $\delta$.  For example, 
\begin{equation}
[\xi_Z]_{WY} = -e^2 \left[ 1 + \frac{\alpha S}{4 s^2 c^2}\right]
\end{equation}
By comparing the residues as written in this way with the expressions we will  
derive for them in terms of the gauge boson spectrum in Sections 5, 6, and 8, we will be able to 
solve for the values of $S$, $T$, $\Delta\rho$ and $\delta$ in terms of the $\Sigma_i$.  

In order to make contact with experiment, we will need to use the more precisely measured $G_F$ 
as an input instead of $M_W$.  In this language, the standard-model 
weak mixing angle $s_Z$ 
is defined as 
\begin{equation}
  c_Z^2 ( 1 - c_Z^2) \equiv \dfrac{\pi \alpha}{\sqrt{2} G_F M_Z^2}, \qquad
  s_Z^2 \equiv 1 - c_Z^2.
\label{eq:Z_scheme}
\end{equation}
This relationship is altered
by the non-standard elements of our model;  thus 
the weak mixing angle appearing in the amplitude for four-fermion processes
is shifted from the standard model value by an amount $\Delta_Z$
\begin{equation}
  c^2 = c_Z^2 + \Delta_Z.
\label{eq:Delta_Z}
\end{equation}
The size of $M_W$ relative to $M_Z$ will likewise be altered from its standard model value.

Deducing the value of $M_Z^2$ from the location of the $Q^2 = - M_Z^2$ pole of (\ref{eq:NC-M}),and comparing this with equations (\ref{eq:Z_scheme}) and (\ref{eq:Delta_Z}) yields the shift in the weak mixing angle:
\begin{equation}
  \Delta_Z = \dfrac{\alpha}{c_Z^2 - s_Z^2}\left[ - \frac{1}{4}( S + \delta) 
  + s_Z^2 c_Z^2 T\right].
  \label{eq:64}
\end{equation}
Taking $M_W^2$ from (\ref{eq:CC-M}) and incorporating the shift
(\ref{eq:Delta_Z}) in the weak mixing angle gives:
\begin{equation}
  M_W^2 = c_Z^2 M_Z^2 
  \left[1 + \dfrac{\alpha}{c_Z^2 - s_Z^2} \left(
         -\frac{1}{2} S + c_Z^2 T - \dfrac{\delta}{4 c_Z^2}
        \right)
  \right].
\label{eq:Z_scheme_MW}
\end{equation}

We can now rewrite the residues in a consistent language in terms of $\alpha$, $G_F$ and $M_Z$.
We use (\ref{eq:Delta_Z}) and (\ref{eq:64}) to rewrite the residues as deduced from  (\ref{eq:NC-M}) and (\ref{eq:CC-M}) in terms of $c_Z$:
\begin{eqnarray}
   \dfrac{1}{e^2}[\xi_Z]_{WW}
  &=& \dfrac{c_Z^2}{s_Z^2} 
     \left[ 1 + \dfrac{1}{c_Z^2 s_Z^2}\left(\Delta_Z + \dfrac{\alpha S}{4}\right)\right]
    \label{eq:Z_scheme_WW}\\ 
    \dfrac{1}{e^2}[\xi_Z]_{WY}
  &=& 
    -1 -  \dfrac{\alpha}{4s_Z^2 c_Z^2} S.
\label{eq:Z_scheme_WY} \\
  \dfrac{1}{e^2}[\xi_Z]_{YY}
  &=& \dfrac{s_Z^2}{c_Z^2} 
     \left[ 1 + \dfrac{1}{c_Z^2 s_Z^2}\left(- \Delta_Z + \dfrac{\alpha S}{4}\right)\right]
    \label{eq:Z_scheme_YY}\\
      \dfrac{1}{e^2}[\xi_W]_{WW}
  &=& \dfrac{1}{s_Z^2} 
     \left[ 1 + \dfrac{1}{s_Z^2}\left(\Delta_Z + \dfrac{\alpha S}{4}\right)\right]
    \label{eq:W_scheme_WW}
\end{eqnarray}
and with $\Delta_Z$ as defined by equation (\ref{eq:Delta_Z}).

Before calculating explicit expressions for all of the pole residues in terms of the sums over mass eigenvalues $\Sigma_i$, we can extract a few more pieces of information from the form of the 
weak amplitudes (\ref{eq:NC-M}) and (\ref{eq:CC-M}).  Evaluating the matrix elements at $Q^2=0$, we find
\begin{equation}
[G_{\rm CC}(0)]_{WW}=4 \sqrt{2} G_F = \dfrac{[\xi_W]_{WW}}{M_W^2}
   +\sum_{w=1}^{N} \dfrac{[\xi_{Ww}]_{WW}}{ {\mathsf m}_{Ww}^2}~,
\label{eq:gf_delta}
\end{equation}
and by removing the pole terms corresponding to the photon, $W$,
and $Z$, we find
\begin{eqnarray}
4\sqrt{2}G_F(\Delta \rho - \alpha T)
& = & \sum_{z=1}^{K} \dfrac{[\xi_{Zz}]_{YY}}{{\mathsf m}_{Zz}^2}~, \\
\sqrt{2} G_F {\alpha \delta \over s^2 c^2} & = &
\sum_{w=1}^{N} \dfrac{[\xi_{Ww}]_{WW}}{ {\mathsf m}_{Ww}^2} = 
\sum_{z=1}^{K} \dfrac{[\xi_{Zz}]_{WW}}{ {\mathsf m}_{Zz}^2}~.
\label{eq:secono}
\end{eqnarray}
Note also that for the amplitudes in equations (\ref{eq:NC-M}) and (\ref{eq:CC-M}) to be consistent, we must have
\begin{equation}
\sum_{z=1}^{K} \dfrac{[\xi_{Zz}]_{WY}}{ {\mathsf m}_{Zz}^2} = 0 
\label{eq:new26}
\end{equation}
to this order.  We will apply these findings in Sections 6 and 8.


\section{Correlation Function \protect{$\bf [G_{\rm NC}(Q^2)]_{WY}$} in a General Linear Moose}

In this section, we will find  $[G_{\rm NC}(Q^2)]_{WY}$ for the general case with both $p$ and $q$ left arbitrary.  We will discuss the other correlation functions and their phenomenology in subsequent sections.

We start with the weak-hypercharge interference term in equation (\ref{nclagrangian}).
Direct calculation of the matrix inverse involved in $D^Z_{p,q}(Q^2)$ involves
the computation of the cofactor related to the $(p,q)$ element of the
matrix $Q^2 {\cal I} +M^2_{Z}$.  Inspection of the neutral-boson mass matrix
eqn. (\ref{eq:neutralmatrix}) and subsequent definitions shows that the cofactor 
is the determinant of a matrix whose
upper $p \times p$ block involves ${\cal{M}}^2_p$, and whose lower $(K-q+1) \times (K-q+1)$ block involves ${\cal{M}}^2_q$.  The middle block is upper-diagonal and
has diagonal entries independent of $Q^2$ and its determinant is therefore constant.  Accordingly, we have 
\begin{equation}
  [G_{\rm NC}(Q^2)]_{WY} = 
  \dfrac{C\,\det[{\cal D}^{p}(Q^2)]^{-1}\det[{\cal D}^{q}(Q^2)]^{-1} }{\det[D^{Z}(Q^2)]^{-1}}
  = 
  \dfrac{C\,\prod_{\hat{p}=1}^{p} (Q^2 + {\mathsf m}_{\hat{p}}^2)
  \prod_{\hat{q}=q+1}^{K+1} (Q^2 + {\mathsf m}_{\hat{q}}^2)}
        {Q^2 (Q^2 + M_Z^2) 
         \prod_{{z}=1}^{K} (Q^2 + {\mathsf m}_{Z{z}}^2)}.
\end{equation}
where $C$ is a constant.  Requiring the residue of the photon pole at  $Q^2=0$ to equal $e^2$ determines the value of $C$ and reveals
\begin{equation}
  [G_{\rm NC}(Q^2)]_{WY} =  \frac{e^2 M_Z^2}{Q^2 (Q^2 + M_Z^2)} 
\left[\prod_{{z}=1}^K \frac{{\mathsf m}_{Z{z}}^2}{Q^2 + {\mathsf m}^2_{Z{z}}} \right]
\left[\prod_{\hat{p}=1}^p \frac{Q^2 + {\mathsf m}^2_{\hat{p}}} {{\mathsf m}_{\hat{p}}^2}\right]
\left[\prod_{\hat{q}=q+1}^{K+1} \frac{Q^2 + {\mathsf m}^2_{\hat{q}}} {{\mathsf m}_{\hat{q}}^2}\right]
\label{gncwy}
\end{equation}

We can read off the Z-pole of the correlation function to find
\begin{equation}
[\xi_Z]_{WY} = - {e^2}
\left[\prod_{{z}=1}^K \frac{{\mathsf m}_{Z{z}}^2}{{\mathsf m}^2_{Z{z}}-M_Z^2} \right]
\left[\prod_{\hat{p}=1}^p \frac{{\mathsf m}^2_{\hat{p}}-M_Z^2} {{\mathsf m}_{\hat{p}}^2}\right]
\left[\prod_{\hat{q}=q+1}^{K+1} \frac{{\mathsf m}^2_{\hat{q}}-M_Z^2} {{\mathsf m}_{\hat{q}}^2}\right]
\label{eq:Z_residue_WY}
\end{equation}
In the limit ${\mathsf m}_{Z,{z>0}},\, {\mathsf m}_{\hat{p}},\, {\mathsf m}_{\hat{q}} \to \infty$, $[\xi_{Z}]_{WY}
\to -e^2$ as at tree-level in the standard model.  Expanding the products in equation (\ref{eq:Z_residue_WY}) to first non-trivial order in the ratio of $M_Z^2/M^2$ (where $M$ is any of $[{\mathsf m}_{Z,{z>0}}, {\mathsf m}_p, {\mathsf m}_q]$, we can rewrite this in shorthand as
\begin{equation}
[\xi_Z]_{WY} = -e^2 [ 1 + M_Z^2 (\Sigma_Z  - \Sigma_p - \Sigma_{q} )]
\label{eq:xiz-wy}
\end{equation}
where the $\Sigma_i$ are defined in (\ref{eq:sig-def}, \ref{eq:sig-def-a}).
 If we set $p=0$ and $q=N+1$ we appropriately recover the leading-order value of $[\xi_Z]_{WY}$ in \cite{Chivukula:2004pk}. From eqn. (\ref{eq:Z_scheme_WY}), this immediately leads to a value for
 the $S$ parameter
\begin{equation}
\alpha S = 4 s_Z^2 c_Z^2 M_Z^2 (\Sigma_Z - \Sigma_p - \Sigma_q)~.
\label{eq:SS1}
\end{equation}

We can also read off the heavy $Z$ pole residues 
\begin{equation}
  [\xi_{Z{k}}]_{WY} =
    e^2 \dfrac{M_Z^2}{{\mathsf m}_{Z{k}}^2- M_Z^2}
    \left[\prod_{{z}\ne {k}}
         \dfrac{{\mathsf m}_{Z{z}}^2}
               {{\mathsf m}_{Z{z}}^2  - {\mathsf m}_{Z{k}}^2 }
    \right]
    \left[\prod_{\hat{p}=1}^{p} 
         \dfrac{{\mathsf m}_{\hat{p}}^2 - {\mathsf m}_{Z{k}}^2}
               {{\mathsf m}_{\hat{p}}^2}
    \right]
    \left[\prod_{\hat{q}=q+1}^{K+1} 
         \dfrac{{\mathsf m}_{\hat{q}}^2 - {\mathsf m}_{Z{k}}^2}
               {{\mathsf m}_{\hat{q}}^2}
    \right].
\label{eq:Z_ell_residue_WY}
\end{equation}
Heavy boson exchange could make significant contributions to the four-fermion interaction processes between a weak current and a hypercharge current only if these residues were appreciable.  Note that $ [\xi_{Zk}]_{WY}$ is, in fact, of order ${\cal O}(M_{Z,W}^2/{\mathsf m}^2_{Zz,Ww})$, and therefore equation (\ref{eq:new26}) holds to this order.  We will gather the heavy pole residues in Appendix C, since they will not be part of the main argument of the paper.


\section{Case I:  \protect{$q = N+1$} \\ Correlation Functions and Consistency Relations}

In this section, we consider the phenomenologically important special case (dubbed Case I) of the general linear moose in which $p$ is left free but $q$ is fixed as $q=N+1$.   We start by writing $[G_{\rm NC}(Q^2)]_{WY}$ for Case I using equation (\ref{gncwy}) above.  We then note that 
$[G_{\rm NC}(Q^2)]_{YY}$ is the same in Case I as in the model of ref. \cite{Chivukula:2004pk} because the latter model also has $q=N+1$ (this correlation function depends on $q$ but not on $p$).  We next calculate $[G_{\rm NC}(Q^2)]_{WW}$ and $[G_{\rm CC}(Q^2)]_{WW}$ directly in terms of eigenvalues of submooses of the full linear moose.  For comparison we then use $[\xi_Z]_{YY}$, $[\xi_Z]_{WY}$ and the consistency conditions (\ref{consistency}) to deduce another way of writing $[G_{NC}(Q^2)]_{WW} - [G_{\rm CC}(Q^2)]_{WW}$ for Case I.  This allows us to derive a relationship among Case I correlation functions
\begin{equation}
[G_{\rm NC}(Q^2) - G_{\rm CC}(Q^2)]_{WW} [G_{\rm NC}(Q^2)]_{YY} = [G_{\rm NC}(Q^2)]_{WY}^2
\label{eq:corr-case1}
\end{equation}
that echos the consistency conditions (\ref{consistency})  for the pole residues.  This relationship was also noted in the specific Case I model discussed in \cite{Chivukula:2004pk}. 

\subsection{ \protect{$\bf [G_{NC}(Q^2)]_{WY}$}}

This correlation function is as computed in Section 5.1, with $q$ taking the value $N+1$.  We immediately see that equation (\ref{eq:xiz-wy}) takes the form
\begin{equation}
[\xi_Z]_{WY} = -e^2 [ 1 + M_Z^2 (\Sigma_Z  - \Sigma_p - \Sigma_{M} )]
\label{eq:fourno}
\end{equation}

\subsection{ \protect{$\bf [G_{NC}(Q^2)]_{YY}$}}

Since this correlation function depends on $q$ but not $p$, it is the same as 
in \cite{Chivukula:2004pk}.
Direct calculation of the matrix inverse involved in $D^Z_{q,q}(Q^2)$ involves
the computation of the cofactor related to the $(N+1,N+1)$ element of the
matrix [$Q^2 {\cal I} +M^2_Z$]. Inspection of the neutral-boson mass matrix eqn. (\ref{eq:mass_matrixN2}) leads to
\begin{eqnarray}
  [D^Z(Q^2)]_{N+1,N+1} 
  &=& 
  \dfrac{\det[D^{W}(Q^2)]^{-1} \det[{\cal D}^{M}(Q^2)]^{-1}}
        {\det[D^{Z}(Q^2)]^{-1}}
  \nonumber\\
  &=&
  \dfrac{(Q^2 + M_W^2)
         \prod_{{w}=1}^N (Q^2 + {\mathsf m}_{W{w}}^2)
         \prod_{{m}=1}^M (Q^2 + {\mathsf m}_{{m}}^2)
        }
        {Q^2 (Q^2 + M_Z^2)
         \prod_{{z}=1}^K (Q^2 + {\mathsf m}_{Z{z}}^2).
        }.
\label{eq:yy_26}
\end{eqnarray}
As noted earlier, charge universality for the photon tells us that the residue of $[G_{\rm NC}(Q^2)]_{YY}$ at $Q^2=0$ is $e^2$, so that
\begin{equation}
  [G_{\rm NC}(Q^2)]_{YY}
  = \dfrac{e^2}{Q^2}\dfrac{[Q^2+M_W^2]M_Z^2}{M_W^2[Q^2+M_Z^2]}
    \left[
      \prod_{{w}=1}^{N}
      \dfrac{Q^2+{\mathsf m}_{W{w}}^2}{{\mathsf m}_{W{w}}^2}
    \right]
     \left[
      \prod_{{z}=1}^{K}
      \dfrac{{\mathsf m}_{Z{z}}^2}{Q^2+{\mathsf m}_{Z{z}}^2}
    \right]
    \left[
      \prod_{{m}=1}^{M}
      \dfrac{Q^2+{\mathsf m}_{{m}}^2}{{\mathsf m}_{{m}}^2}
    \right].
\label{eq:exact_YY}
\end{equation}
Reading off the residue of the Z pole in Eq.(\ref{eq:exact_YY}), we obtain
\begin{equation}
  [\xi_Z]_{YY} 
  = e^2 \dfrac{M_Z^2- M_W^2}{M_W^2}
      \left[\prod_{{w}=1}^N
      \dfrac{{\mathsf m}_{W{w}}^2-M_Z^2}{{\mathsf m}_{W{w}}^2}
      \right] 
      \left[\prod_{{z}=1}^K
      \dfrac{{\mathsf m}_{Z{z}}^2}{{\mathsf m}_{Z{z}}^2-M_Z^2}
      \right]
      \left[\prod_{{m}=1}^M
      \dfrac{{\mathsf m}_{{m}}^2-M_Z^2}{{\mathsf m}_{{m}}^2}
      \right],
\label{eq:Z_residue_YY}
 \end{equation}
Note that, in the limit ${\mathsf m}_{Z,{z>0}}, {\mathsf m}_{W,{w>0}}, {\mathsf m}_{{m}} \to 
\infty$, $[\xi_Z]_{YY} \to e^2 (M^2_Z - M^2_W)/M^2_W$, so that the correct standard
model tree-level value of the $Z$-boson coupling to hypercharge is recovered.
If we expand the expression for $[\xi_Z]_{YY}$ to first non-trivial order, we can rewrite it  as
\begin{equation}
[\xi_Z]_{YY} = \frac{e^2 (M_Z^2 - M_W^2)}{M_W^2} [1 + M_Z^2 (\Sigma_Z - \Sigma_W - \Sigma_{M})]
\label{eq:exp_xi_Z_YY}
\end{equation}

\subsection{\protect{$\bf [G_{NC}(Q^2)]_{WW}$ and $\bf [G_{CC}(Q^2)]_{WW}$}}

In terms of the matrices
defined previously, we may calculate the neutral-current correlation function directly 
(in terms of the relevant cofactors) as
\begin{equation}
[G_{\rm NC}(Q^2)]_{WW}  =  {e^2 M^2_Z \over Q^2  [Q^2+M^2_Z]}{[Q^2+M^2_L] \over M^2_L}
\left[\prod_{z=1}^K {{\mathsf m}^2_{Zz}\over Q^2+{\mathsf m}^2_{Zz}}\right]
\left[\prod_{\hat{p}=1}^p {Q^2+{\mathsf m}^2_{\hat{p}} \over {\mathsf m}^2_{\hat{p}}}\right]
\left[\prod_{{l}=1}^{K-p} {Q^2+{\mathsf m}^2_l \over {\mathsf m}^2_l}\right]~,
\label{eq:exact_gnc_ww}
\end{equation}
where we have displayed the contribution corresponding to the light eigenvalue $M^2_L$ of the
matrix ${\cal M}^2_L$ explicitly, and charge universality determines the value at $Q^2=0$.
Similarlly, the charged-current correlation function may be written as
\begin{equation}
[G_{\rm CC}(Q^2)]_{WW} =  {4 \sqrt{2} G_F M^2_W \over [Q^2+M^2_W]}\
\left[\prod_{w=1}^N {{\mathsf m}^2_{Ww}\over Q^2+{\mathsf m}^2_{Ww}}\right]
\left[\prod_{\hat{p}=1}^p {Q^2+{\mathsf m}^2_{\hat{p}} \over {\mathsf m}^2_{\hat{p}}}\right]
\left[\prod_{r=p+1}^{N} {Q^2+{\mathsf m}^2_r \over  {\mathsf m}^2_r} \right]~,
\label{eq:exact_gcc_ww}
\end{equation}
where we have imposed the relation $[G_{CC}(0)]_{WW} = 4 \sqrt{2} G_F$.

Reading off the residues of the poles at $M^2_{Z}$ and $M^2_{W}$, we find
\begin{eqnarray}
[\xi_W]_{WW} & = & {4\sqrt{2} G_F M^2_W}
\left[ \prod_{w=1}^N {{\mathsf m}^2_{Ww} \over  {\mathsf m}^2_{Ww} - M^2_W}\right]
\left[ \prod_{\hat{p}=1}^p {{\mathsf m}^2_{\hat{p}} -M^2_W \over  {\mathsf m}^2_{\hat{p}}}\right]
\left[ \prod_{r=p+1}^{N} {{\mathsf m}^2_{r} -M^2_W \over  {\mathsf m}^2_{r}}\right]~,
\label{eq:W_residue_WW_new} \\
{}
[\xi_Z]_{WW} & = & -e^2\,{[M^2_L - M^2_Z]\over M^2_L}
\left[ \prod_{z=1}^K {{\mathsf m}^2_{Zz} \over  {\mathsf m}^2_{Zz} - M^2_Z}\right]
\left[ \prod_{\hat{p}=1}^p {{\mathsf m}^2_{\hat{p}} -M^2_Z \over  {\mathsf m}^2_{\hat{p}}}\right]
\left[ \prod_{l=1}^{K-p} {{\mathsf m}^2_{l} -M^2_Z \over  {\mathsf m}^2_{l}}\right]~.
\label{eq:Z_residue_WW_new}
\end{eqnarray}
Expanding $[\xi_{W,Z}]_{WW}$ to lowest non-trivial order, we find
\begin{eqnarray}
[\xi_W]_{WW} & = & 4 \sqrt{2} G_F M^2_W \left(1+M^2_W(\Sigma_W - \Sigma_p -\Sigma_r)\right)~,
\label{eq:firsno_new} \\
{}
[\xi_Z]_{WW} & = & -e^2\,\left[{M^2_L - M^2_Z \over M^2_L}\right]
\left(1 + M^2_Z(\Sigma_Z-\Sigma_p-\Sigma_L)\right)~.
\label{eq:thirno_new}
\end{eqnarray}

\subsection{ \protect{$\bf [G_{NC}(Q^2)]_{WW} - [G_{CC}(Q^2)]_{WW}$}: Consistency Relations}

\EPSFIGURE[t]{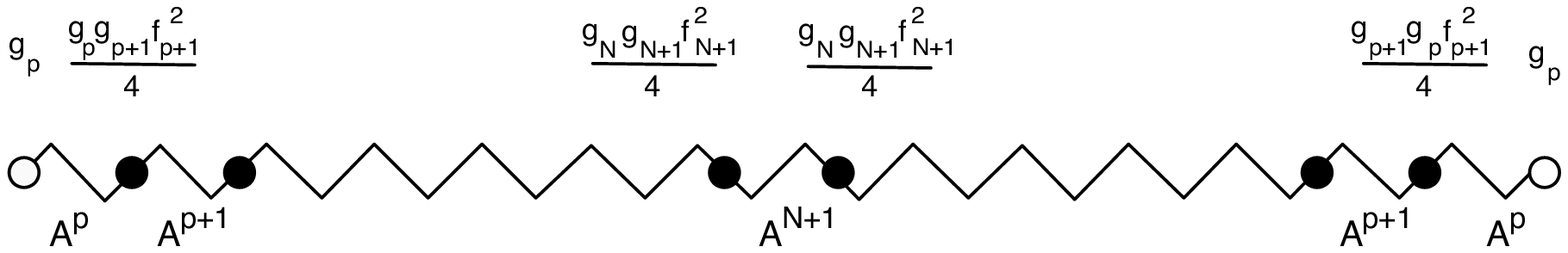,width=0.9\textwidth}
{Leading diagram at high-$Q^2$ ($Q^2\gg {\mathsf m}^2_{Zz,Ww}$
for $w,z>0$) which 
distinguishes $[G_{NC}(Q^2)]_{WW}$ from
$[G_{CC}(Q^2)]_{WW}$.
\label{fig:three}}

Next we turn to examining the difference
\begin{equation}
\label{eq:NC_CC_WW}
  [G_{\rm NC}(Q^2)]_{WW} - [G_{\rm CC}(Q^2)]_{WW} \,.
\end{equation}
In the \,$g_{N+1}^{~}\to 0$\, limit, 
the matrix $M_Z^2$ in (\ref{eq:mass_matrixN2}) becomes block-diagonal
and the $U(1)$-sector fully decouples, leading to a vanishing
difference in Eq.\,(\ref{eq:NC_CC_WW}).
As the custodial violation \,$g_{N+1}\neq 0$\, originates from
the $(N+1)$th site of the moose chain,  at high-$Q^2$ the asymptotic
behavior of (\ref{eq:NC_CC_WW})  arises from the
diagram shown in Fig. \ref{fig:three} (c.f. \cite{Chivukula:2004pk})
\begin{equation}
  [G_{\rm NC}(Q^2)]_{WW} - [G_{\rm CC}(Q^2)]_{WW}
  ~\propto~ \dfrac{g_{N+1}^2}{~(Q^2)^{2(N-p)+3}~} .
\label{eq:Asym_NC_CC_WW}
\end{equation}

Using equations (\ref{eq:NC_WW}) and (\ref{eq:CC_WW}), we see that
the only consistent way to achieve the high energy behavior 
(\ref{eq:Asym_NC_CC_WW}) is
\begin{equation}
  [G_{\rm NC}(Q^2)]_{WW}  - [G_{\rm CC}(Q^2)]_{WW} 
= \dfrac{e^2 M_W^2 M_Z^2 \,{\cal R}(Q^2)}{Q^2[Q^2+M_W^2][Q^2 + M_Z^2]}
        \left[
        \prod_{{w}=1}^{N} 
        \dfrac{{\mathsf m}_{W{w}}^2}
              {Q^2 + {\mathsf m}_{W{w}}^2}
        \right]
         \left[
        \prod_{{z}=1}^{K} 
        \dfrac{{\mathsf m}_{Z {z}}^2}
              {Q^2 + {\mathsf m}_{Z{z}}^2}
        \right]~,
\label{eq:exact_WW1}
\end{equation}
where the function ${\cal R}(Q^2)$ is a polynomial in $Q^2$ of order $M + 2p$ and ${\cal  R}(0) = 1$ in order to have (\ref{eq:exact_WW1}) satisfy charge universality.  

From eqns. (\ref{eq:exact_gnc_ww}) and (\ref{eq:exact_gcc_ww}), we see that the polynomial
${\cal R}(Q^2)$, must vanish at $Q^2=-{\mathsf m}^2_{\hat{p}}$  -- yielding
$p$ conditions. Furthermore, the residues $[\xi_Z]_{WW}$ and $[\xi_{Zk}]_{WW}$ 
must satisfy the $K+1$ consistency conditions of eqn. (\ref{consistency}). Together, these yield
$K+p+1$ conditions, which is greater than the  $M+2p-1$ free parameters in the
polynomial ${\cal R}(Q^2)$. Therefore the polynomial ${\cal R}(Q^2)$ is
(over)-constrained by these conditions.  Motivated by the relation eqn. (\ref{eq:corr-case1})
found in \cite{Chivukula:2004pk}, and recalling that $M^2_q \equiv {\cal M}^2_M$
in this case, we consider the ansatz
\begin{equation}
{\cal R}(Q^2) = \left[\prod_{m=1}^M{Q^2+{\mathsf m}^2_m \over {\mathsf m}^2_m}\right]
\left[\prod_{\hat{p}=1}^p {Q^2+{\mathsf m}^2_{\hat{p}} \over {\mathsf m}^2_{\hat{p}}}\right]^2~,
\end{equation}
and find that the corresponding difference of correlation functions
\begin{eqnarray}
  [G_{\rm NC}(Q^2)]_{WW}  &-& [G_{\rm CC}(Q^2)]_{WW} 
= \dfrac{e^2 M_W^2 M_Z^2}
           {~Q^2[Q^2+M_W^2][Q^2 + M_Z^2]~} \label{eq:exact_WWff}
\\
  &\times& \left[\prod_{{w} =1}^{N} 
        \dfrac{{\mathsf m}_{W{w}}^2}
              {\,Q^2 + {\mathsf m}_{W{w}}^2\,} \right]
      \,\left[\prod_{{z} =1}^{K} 
        \dfrac{{\mathsf m}_{Z {z}}^2}
              {\,Q^2 + {\mathsf m}_{Z {z}}^2\,}\right]
      \,\left[\prod_{{m} =1}^{M} 
        \dfrac{\,Q^2 + {\mathsf m}_{{m}}^2\,}{{\mathsf m}_{{m}}^2}\right]
      \,\left[\prod_{\hat{p} =1}^{p} 
        \dfrac{\,(Q^2+{\mathsf m}_{\hat{p}}^2)^2 \, , } 
              {{\mathsf m}_{\hat{p}}^4} \right]\nonumber
\end{eqnarray}
satisfies all of the consistency conditions. Therefore our ansatz for
${\cal R}(Q^2)$ is correct. Explicit examination of equations (\ref{gncwy}), (\ref{eq:exact_YY}) and (\ref{eq:exact_WWff})  reveals the  promised relationship (\ref{eq:corr-case1}).

The pole residues are
\begin{equation}
  {[\xi_W^{~}]}_{WW}^{~} = 
   \dfrac{e^2 M_Z^2}{\,M_Z^2 \!-\! M_W^2\,}
   \left[\prod_{{w} =1}^{N} 
     \dfrac{{\mathsf m}_{W{w}}^2}
           {\,{\mathsf m}_{W{w}}^2 \!-\! M_W^2\,}\right]
   \left[\prod_{{z} =1}^{K} 
     \dfrac{{\mathsf m}_{Z{z}}^2}
           {{\mathsf m}_{Z{z}}^2 \!-\! M_W^2\,}\right]
   \left[\prod_{{m} =1}^{M} 
     \dfrac{\,{\mathsf m}_{{m}}^2 - M_W^2\,}{{\mathsf m}_{{m}}^2}\right]
   \left[\prod_{\hat{p} =1}^{p} 
     \dfrac{\,({\mathsf m}_{\hat{p}}^2 - M_W^2)^2\,}
           {{\mathsf m}_{\hat{p}}^4} \right]\,,
\label{eq:W_residue_WW}
\end{equation}
\begin{equation}
  {[\xi_Z^{~}]}_{WW}^{~} =
   \dfrac{e^2 M_W^2}{\,M_Z^2 \!-\! M_W^2\,}
   \left[\prod_{{w} =1}^{N} 
     \dfrac{{\mathsf m}_{W{w}}^2}
           {\,{\mathsf m}_{W{w}}^2 \!-\! M_Z^2\,}\right]
   \left[\prod_{{z} =1}^{K} 
     \dfrac{{\mathsf m}_{Z{z}}^2}
           {{\mathsf m}_{Z{z}}^2 \!-\! M_Z^2\,}\right]
   \left[\prod_{{m} =1}^{M} 
     \dfrac{\,{\mathsf m}_{{m}}^2 - M_Z^2\,}{{\mathsf m}_{{m}}^2}\right]
   \left[\prod_{\hat{p} =1}^{p} 
     \dfrac{\,({\mathsf m}_{\hat{p}}^2 - M_Z^2)^2\,}
           {{\mathsf m}_{\hat{p}}^4}\right] \,,
\label{eq:Z_residue_WW}
\end{equation}
If we expand the expressions for $[\xi_Z]_{WW}$ and $[\xi_Z]_{WW}$ to first non-trivial order, we can rewrite them  as
\begin{equation}
[\xi_W]_{WW} = \frac{e^2 M_Z^2}{M_Z^2 - M_W^2} [ 1 + M_W^2 (\Sigma_Z  + \Sigma_W - \Sigma_{M} - 2\Sigma_{p})]
\label{eq:firsno}
\end{equation}
\begin{equation}
[\xi_Z]_{WW} = \frac{e^2 M_W^2}{M_Z^2 - M_W^2} [ 1 + M_Z^2 (\Sigma_Z  + \Sigma_W - \Sigma_{ M} - 2\Sigma_{p})]
\label{eq:thirno}
\end{equation}
 If we set $p=0$ we recover our leading-order values from \cite{Chivukula:2004pk}.
 
 Comparing eqns. (\ref{eq:firsno_new}) and (\ref{eq:firsno}), we find the relation
\begin{equation}
{e^2 \over 4\sqrt{2} G_F} = M^2_W \left(1-{M^2_W \over M^2_Z}\right)
[1-M^2_W(\Sigma_Z+\Sigma_r-\Sigma_p-\Sigma_M)]~,
\label{consistencyvi}
\end{equation}
to this order. By comparing eqns. (\ref{eq:thirno_new}) and (\ref{eq:thirno}), we find
\begin{equation}
M^2_L = (M^2_Z - M^2_W) [1-M^2_W(\Sigma_W + \Sigma_L - \Sigma_M - \Sigma_p)]~,
\label{consistencyvii}
\end{equation}
which demonstrates our assertion that the matrix ${\cal M}^2_L$ has a light eigenvalue.

Using eqns. (\ref{gncwy}), (\ref{eq:exact_YY}), (\ref{eq:exact_gnc_ww}) and
(\ref{eq:exact_gcc_ww}), we see that the consistency relation of
eqn. (\ref{eq:corr-case1}) can be written 
schematically as
\begin{equation}
\prod_{w,l}\left(1+{Q^2\over{\mathsf m}^2_{Ww}}\right)
\left(1+{Q^2\over {\mathsf m}^2_l}\right) =
{\sqrt{2} G_F Q^2 \over e^2} \prod_{r,z}\left(1+{Q^2\over{\mathsf m}^2_{r}}\right)
\left(1+{Q^2\over {\mathsf m}^2_{Zz}}\right) +
\prod_{\hat{p}, \hat{q}}\left(1+{Q^2\over{\mathsf m}^2_{\hat{p}}}\right)
\left(1+{Q^2\over {\mathsf m}^2_{\hat{q}}}\right) ~,
\label{consistency0}
\end{equation}
where the products run over all the corresponding (non-zero) eigenvalues, including
the light eigenvalues ${\mathsf m}^2_{Z0}\equiv M^2_Z$ ,${\mathsf m}^2_{W0} \equiv M^2_W$, 
and ${\mathsf m}^2_{l=0} \equiv M^2_L$.
In principle this expression gives us many relations among the various
eigenvalues. In practice, however, we are only interested in the consequences at momenta
$Q^2$ much less than any of the heavy eigenvalues. Expanding eqn. (\ref{consistency0})
to lowest non-trivial order, we find
\begin{eqnarray}
\left(1+{Q^2\over M^2_L}\right) \left(1 + {Q^2\over M^2_W}\right) 
(1+ Q^2 \Sigma_{ L} + Q^2 \Sigma_W)
& = & {4\sqrt{2} G_F Q^2\over e^2 }  \left(1+{Q^2\over M^2_Z}\right)(1+Q^2 \Sigma_r + Q^2 \Sigma_Z)
\nonumber \\
& + & (1+Q^2 \Sigma_p + Q^2 \Sigma_q)~,
\label{consistency1}
\end{eqnarray}
where we have explicitly kept all terms involving only light masses. Equating the
terms proportional to $Q^6$, $Q^4$, and $Q^2$ (consistently to this order in
the heavy mass expansion)  we find
\begin{equation}
\Sigma_{ L} + \Sigma_W = \Sigma_r + \Sigma_Z~,
\label{consistencyiii}
\end{equation}
and reproduce eqns. (\ref{consistencyvi}) and (\ref{consistencyvii}).

\section{Case I: \protect{$q=N+1$} \\
Electroweak Phenomenology}

\subsection{$\Delta \rho$}

Because Case I models satisfy the consistency relation of
eqn. (\ref{eq:corr-case1}), the low-energy $\rho$ parameter will
always equal 1 in this class of models -- just as was found for the specific Case I model of \cite{Chivukula:2004pk}. 

In discussing low-energy interactions it is conventional to rewrite  the neutral-current Lagrangian of eqn. (\ref{nclagrangian}) in terms of weak and electromagnetic currents
as
\begin{equation}
{\cal L}_{nc} = - {1\over 2} A(Q^2)  J^\mu_3 J_{3\mu} - B(Q^2) J^\mu_3 J_{Q\mu}
- {1\over 2} C(Q^2) J^\mu_Q J_{Q\mu}~,
\end{equation}
where
\begin{eqnarray}
A(Q^2) & = & [G_{\rm NC}(Q^2)]_{WW} - 2 [G_{\rm NC}(Q^2)]_{WY}+ [G_{\rm NC}(Q^2)]_{YY} \\
B(Q^2) & = & [G_{\rm NC}(Q^2)]_{WY} - [G_{\rm NC}(Q^2)]_{YY} \\
C(Q^2) & = & [G_{\rm NC}(Q^2)]_{YY}~.
\end{eqnarray}
As required, the photon pole cancels in the expressions for $A(Q^2)$ and
$B(Q^2)$.
The low-energy $\rho$ parameter is defined by
\begin{equation}
\rho = \lim_{Q^2\to 0} {A(Q^2) \over [G_{\rm CC}(Q^2)]_{WW}}~.
\end{equation}
Consider
\begin{equation}
{A(Q^2) \over [G_{\rm CC}(Q^2)]_{WW}} - 1 =
{([G_{\rm NC}(Q^2)]_{WY}-[G_{\rm NC}(Q^2)]_{YY})^2 \over [G_{\rm NC}(Q^2)]_{YY} [G_{\rm CC}(Q^2)]_{WW}}~,
\end{equation}
where the equality holds because of eqn. (\ref{eq:corr-case1}). As $Q^2 \to 0$ in this
expression, charge universality insures that the photon pole contribution cancels
in the numerator, while the denominator diverges like $e^2/Q^2$. Therefore,
\begin{equation}
\Delta \rho = \lim_{Q^2\to 0} {A(Q^2) \over [G_{\rm CC}(Q^2)]_{WW}} - 1 \equiv 0~.
\end{equation}

\subsection{$\alpha S$, $\alpha T$, and $\alpha \delta$}

To extract the on-shell electroweak parameters from our expressions for the residues of the poles in the correlations functions, we must first
shift to a scheme where $G_F$ is used as an input instead of $M_W$ by applying equation (\ref{eq:Z_scheme_MW}) to equations (\ref{eq:fourno}), (\ref{eq:exp_xi_Z_YY}), (\ref{eq:firsno}) and (\ref{eq:thirno}).  Then comparing $[\xi_Z]_{YY}$, $[\xi_Z]_{WY}$, $[\xi_Z]_{WW}$, and $[\xi_W]_{WW}$ written in terms of the $\Sigma_i$ to the forms derived earlier as (\ref{eq:Z_scheme_WW}) - (\ref{eq:W_scheme_WW}),  we can solve for $S$ and $T$ to leading order:
\begin{equation}
\alpha S = 4 s_Z^2 c_Z^2 M_Z^2 (\Sigma_Z - \Sigma_p - \Sigma_{ M})
\label{eq:SSi}
\end{equation}
\begin{equation}
\alpha T = s_Z^2 M_Z^2 (\Sigma_Z - \Sigma_W  - \Sigma_{ M})
\label{eq:TTi}
\end{equation}
From eqns. (\ref{eq:gf_delta}) and (\ref{eq:secono}) we find
\begin{equation}
4\sqrt{2} G_F =   \dfrac{[\xi_W]_{WW}}{M_W^2}
   + \sqrt{2} G_F {\alpha \delta \over s^2 c^2}~.
   \label{eq:nnb}
\end{equation}
Applying (\ref{eq:firsno_new}) to the RHS of
(\ref{eq:nnb}) we find
\begin{equation}
\alpha \delta  = - 4 s^2_Z c^4_Z M^2_Z (\Sigma_W - \Sigma_r -\Sigma_p)~,
\label{eq:deltai}
\end{equation}
where we have substituted $c_Z^2 M_Z^2$ for $M_W^2$ and 
$s^2_Z$ for $s^2$ to this order since the $\Sigma_i$ and $\delta$ are small quantities.

\subsection{Zero-Momentum Parameters}

Barbieri {\it et al.} have introduced \cite{Barbieri:2004qk} a set of electroweak parameters defined at $Q^2 = 0$ that describe leading-order and higher-order effects of physics beyond the standard model.  In ref. \cite{Chivukula:2004af}, we not only derived the relationships between the zero-momentum parameters and the on-shell parameters $S$, $T$, $\delta$ and $\Delta \rho$, but also quoted expressions for the zero-momentum parameters in terms of the  $\Sigma_i$.  We now show how the expressions we have obtained for the correlation functions in Section 6 support the results described in \cite{Chivukula:2004af}.

Barbieri {\it et al.} \cite{Barbieri:2004qk} choose parameters to describe four-fermion
electroweak processes using the transverse gauge-boson 
polarization amplitudes. Formally, all such processes can be
summarized in momentum space (at tree-level in the electroweak interactions, having integrated
out all heavy states, and ignoring external fermion masses) by  the neutral current Lagrangian
\begin{equation}
{\cal L}_{nc} = {1\over 2} 
\left(
\begin{array}{cc}
J_{3\mu} & J_{B\mu}
\end{array}
\right)
\left[
\begin{array}{cc}
{\Pi}_{W^3 W^3}(Q^2) & {\Pi}_{W^3 B}(Q^2) \\
{\Pi}_{W^3 B}(Q^2) & {\Pi}_{BB}(Q^2)
\end{array}
\right]^{-1}
\left(
\begin{array}{c}
J^\mu_{3} \\
J_B^\mu
\end{array}
\right)~,
\label{bnclagrangian}
\end{equation}
 and the charged
current Lagrangian
\begin{equation}
{\cal L}_{cc}= {1\over 2} \left[\Pi_{W^+ W^-}(Q^2)\right]^{-1} J^\mu_+ J_{-\mu}~,
\label{bcclagrangian}
\end{equation}
where the $\vec{J}^\mu$ and $J_B^\mu$ are the weak isospin and hypercharge
fermion currents respectively. All two-point correlation functions of fermionic currents
-- and therefore all four-fermion scattering amplitudes at tree-level -- 
can be read off from the appropriate element(s) of the inverse
gauge-boson polarization matrix. 

Comparing eqns. (\ref{nclagrangian}) and (\ref{cclagrangian}) to eqns. (\ref{bnclagrangian})
and (\ref{bcclagrangian}), we find
\begin{equation}
{\bf \Pi}^{-1}(Q^2) = \left[
\begin{array}{cc}
{\Pi}_{W^3 W^3}(Q^2) & {\Pi}_{W^3 B}(Q^2) \\
{\Pi}_{W^3 B}(Q^2) & {\Pi}_{BB}(Q^2)
\end{array}
\right]^{-1}
\equiv
\, - \,
\left[
\begin{array}{cc}
[G_{\rm NC}(Q^2)]_{WW}  &  [G_{\rm NC}(Q^2)]_{WY} \cr
[G_{\rm NC}(Q^2)]_{WY}  &  [G_{\rm NC}(Q^2)]_{YY}
\end{array}
\right]~,
\label{pidefine}
\end{equation}
and
\begin{equation}
\Pi_{W^+ W^-} \equiv {-1 \over [G_{\rm CC}(Q^2)]_{WW}}~.
\label{picdefine}
\end{equation}
Using the consistency relation, eqn. (\ref{eq:corr-case1}), we find
\begin{equation}
{\bf \Pi}(Q^2) =
\left[
\begin{array}{cc}
{-1\over [G_{\rm CC}(Q^2)]_{WW} } & {[G_{\rm NC}(Q^2)]_{WY} \over 
[G_{\rm CC}(Q^2)]_{WW}\, [G_{\rm NC}(Q^2)]_{YY}} \cr
\\
{[G_{\rm NC}(Q^2)]_{WY} \over 
[G_{\rm CC}(Q^2)]_{WW}\, [G_{\rm NC}(Q^2)]_{YY}} & 
{-[G_{\rm NC}(Q^2)]_{WW}\over [G_{\rm CC}(Q^2)]_{WW}\,
[G_{\rm NC}(Q^2)]_{YY}}
\end{array}
\right]~.
\label{pi}
\end{equation}

Barbieri {\it et al.} proceed by defining the (approximate) electroweak couplings
\begin{equation}
{1\over g^2} \equiv \left[d\Pi_{W^+ W^-}(Q^2) \over d(-Q^2)\right]_{Q^2=0}
\ \ \ \ , \ \ \ 
{1\over {g'}^2} \equiv \left[{d\Pi_{BB}(Q^2)\over d(-Q^2)}\right]_{Q^2=0}~,
\label{defgs}
\end{equation}
and the electroweak scale
\begin{equation}
v^2 \equiv -4\,\Pi_{W^+ W^-}(0)  = (\sqrt{2} G_F)^{-1}\approx (246\,{\rm GeV})^2~.
\end{equation}
(Our definition of $v$ differs from that used in ref.
\protect\cite{Barbieri:2004qk} by $\sqrt{2}$.) In terms of the polarization functions and these constants, the authors of \cite{Barbieri:2004qk}
define the parameters
\begin{eqnarray}
\hat{S} & \equiv & g^2 \left[{d\Pi_{W^3 B} (Q^2)\over d(-Q^2)}\right]_{Q^2=0}~, \label{defshat}\\
\hat{T} & \equiv &  {g^2 \over M^2_W}\left(\Pi_{W^3W^3}(0)-\Pi_{W^+ W^-}(0)\right)~,\label{defthat}\\
W & \equiv & {g^2 M^2_W \over 2} \left[{d^2 \Pi_{W^3 W^3} (Q^2)\over d(-Q^2)^2}\right]_{Q^2=0}~,\label{defw}\\
Y & \equiv & {{g'}^2 M^2_W \over 2} \left[{d^2 \Pi_{BB}(Q^2) \over d(-Q^2)^2}\right]_{Q^2=0}~. \label{defy}
\end{eqnarray}
In any non-standard electroweak model in which all of the relevant effects occur 
{\it only} in the correlation functions
of fermionic electroweak gauge currents,\footnote{And not, for example, through extra gauge-bosons or
compositeness operators involving the $B-L$ or weak isosinglet currents.}
the values of these four parameters \cite{Barbieri:2004qk} summarize the leading deviations 
in all four-fermion processes from the standard model predictions.  In addition, the quantities 
\cite{Barbieri:2004qk} 
\begin{eqnarray}
\hat{U} & \equiv & - g^2\left[{d\Pi_{W^3 W^3} \over d(-Q^2)} - {d\Pi_{W^+ W^-} \over d(-Q^2)}\right]_{Q^2=0} \\
\hat{V} & \equiv & {g^2 M^2_W \over 2} \left[{d^2\Pi_{W^3 W^3} \over d(-Q^2)^2} - {d^2\Pi_{W^+ W^-} \over d(-Q^2)^2}\right]_{Q^2=0} \\
X & = & {g g' M^2_W \over 2} \left[{d^2\Pi_{W^3 B} (Q^2)\over d(-Q^2)^2}\right]_{Q^2=0}
\end{eqnarray}
describe higher-order effects.

We now evaluate these expressions for the on-shell parameters in Case I linear moose models.  Initially, we notice that 
\begin{equation}
\Pi_{W^+ W^-}(Q^2) = \Pi_{W^3 W^3}(Q^2)~,
\end{equation}
and therefore the zero-momentum parameters
\begin{equation}
\hat{T}=\hat{U}=\hat{V} =0~,
\end{equation}
vanish identically.

Using the forms of the correlation functions derived previously, and the
identities (\ref{pi}) which we have just derived from the consistency relations, we may compute
$\hat{S}$, $W$, and $Y$ directly in terms of the $\Sigma_i$.  Alternatively, as shown in \cite{Chivukula:2004af}, we may compute the zero-momentum parameters from the on-shell parameters $\alpha S$, $\alpha T$, and $\alpha \delta$, which we have already derived in terms of the $\Sigma_i$ in Section 6. In either case, we find
\begin{eqnarray}
\hat{S} & = & M^2_W\, \Sigma_r \\
W & = & - M^2_W ( \Sigma_W - \Sigma_p - \Sigma_r) \\
Y & = & - M^2_W(\Sigma_Z - \Sigma_W - \Sigma_{ M})~.
\end{eqnarray}
We see that $\hat{S}$ is strictly positive -- we will return to the phenomenological
consequences of this in Section 11.

Finally, computing the parameter $X$ in the same way, we see that it is suppressed by four
powers of the heavy masses and is, as expected, zero to this order \cite{Barbieri:2004qk}.

\section{General Linear Moose Models: \protect{$q \geq N+1$} \\
Correlation Functions and Consistency Conditions}

We now study the correlation functions of the general linear moose.  We have already derived $[G_{NC}(Q^2)]_{WY}$  for the general case in Section 5.  Since in both Case I and the general model the value of $p$ is arbitrary (and the charged currents do not depend on $q$), the expressions for $[G_{NC}(Q^2)]_{WW}$ and $[G_{\rm CC}(Q^2)]_{WW}$ and their residues are the same in the general model as in Case I.  We therefore reapply the consistency conditions  (\ref{consistency}) and deduce the residues of $[G_{\rm NC}(Q^2)]_{YY}$ in the general model.  We then discuss how the consistency relations among correlation functions found for Case I are modified in the general linear moose.

\subsection{\protect{$\bf [G_{NC}(Q^2)]_{WY}$} in the General Case}

The expressions for $ [G_{NC}(Q^2)]_{WY}$ and the associated residues are as derived in section 5.

\subsection{\protect{$\bf [G_{NC}(Q^2)]_{WW}$} and \protect{$\bf [G_{CC}(Q^2)]_{WW}$} in the General Case}

As previously noted, the form of $[G_{NC}(Q^2)]_{WW}$ and $[G_{CC}(Q^2)]_{WW}$ depend only on the identity of the $SU(2)$ group to which fermions couple and not on the $U(1)$ to which they couple.
The expressions for $[G_{NC}(Q^2)]_{WW}]$, $[G_{CC}(Q^2)]_{WW}$ and their residues derived in sections 6.3 and 6.4 for Case I therefore apply equally well in the general case.  Note that there is dependence on the ${\cal M}_m$ (which cover all of the $U(1)$ groups with $j > N+1$) rather than on ${\cal M}_q$.  

\subsection{\protect{$\bf [G_{NC}(Q^2)]_{YY}$} in the General Case: The Related Case I Moose}

We can calculate the pole residues of $[G_{NC}(Q^2)]_{YY}$ in the general case by applying the consistency conditions (\ref{consistency}) to our previous results for $[\xi_Z]_{WY}$ (\ref{eq:Z_residue_WY}) and $[\xi_Z]_{WW}$ (\ref{eq:Z_residue_WW}).  We find
\begin{equation}
[\xi_Z]_{YY} = \frac{e^2 (M_Z^2 - M_W^2)}{M_W^2} 
\left[\prod_{{w} =1}^{N} 
     \dfrac{\,{\mathsf m}_{W{w}}^2 \!-\! M_Z^2\,}{{\mathsf m}_{W{w}}^2}\right]
     \left[\prod_{{z}=1}^K \frac{{\mathsf m}_{Z{z}}^2}{{\mathsf m}^2_{Z{z}}-M_Z^2} \right]
     \left[  \prod_{{m} =1}^{M} 
     \dfrac{{\mathsf m}_{{m}}^2} {\,{\mathsf m}_{{m}}^2 - M_Z^2\,}\right]
     \left[\prod_{\hat{q}=q+1}^{K+1} \frac{({\mathsf m}^2_{\hat{q}}-M_Z^2)^2} 
     {{\mathsf m}_{\hat{q}}^4}\right]~.
     \label{eq:gen-Z-YY}
\end{equation}
This reduces to the Case I (and \cite{Chivukula:2004pk}) expression (\ref{eq:Z_residue_YY})
 if we set $q = N+1$, since the ${\mathsf m}_{\hat{q}}$ then 
 take on the values ${\mathsf m}_{\hat{m}}$.  
To leading order in mass-squared ratios, one has
\begin{equation}
[\xi_Z]_{YY} = \frac{e^2 (M_Z^2 - M_W^2)}{M_W^2} [1 + M_Z^2 (\Sigma_Z - \Sigma_W + \Sigma_M - 2 \Sigma_q)]~.
\label{eq:exp_xi_Z_YY_general}
\end{equation}
Note that for $q = N+1$ we have $\Sigma_q = \Sigma_M$ and we recover the Case I result (\ref{eq:exp_xi_Z_YY}); if we set $p=0$ and $q=N+1$ we recover the result in \cite{Chivukula:2004pk}.

\EPSFIGURE[t]{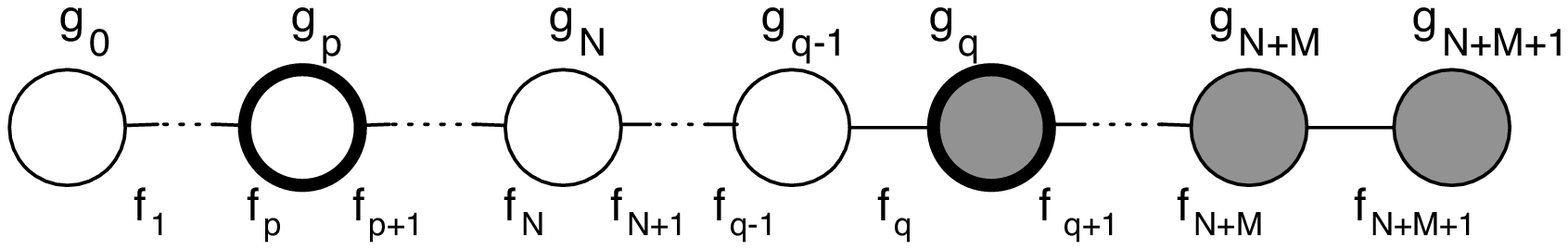,width=0.9\textwidth}
{``Related'' Case I moose corresponding
to the moose in fig. 1. In this moose, the
$SU(2)$ groups continue all the way to site $q-1$.
\label{fig:ttwo}}

Alternatively, we may use a different kind of extrapolation from a Case I moose to
calculate  $[G_{NC}(Q^2)]_{YY}$.  Consider the Case I moose shown in 
fig.\ref{fig:ttwo}: in this model, the values of the couplings and f-constants are 
the same as in our original general moose, but the sites corresponding to $SU(2)$ groups 
now include those all the way to site $q-1$. All of the properties of the neutral
bosons are the same in this ``related Case I moose'' as in our original
moose.  Therefore, we can use our previous knowledge of the form of $[G_{NC}(Q^2)]_{YY}$ 
for Case I to find this correlation function for the moose in fig. \ref{fig:ttwo}, and this will
also be the answer for our original moose.

Furthermore, as we shall see in the next subsection, the properties 
of the charged bosons and the corresponding correlation functions in
the two models are related by a new consistency condition which will allow
us to extract to the electroweak parameters more easily. Defining 
the matrix and  eigenvalues for the charged-boson masses in the 
related Case I moose
\begin{equation}
M^2_{W'}  =  M^2_{[0,q)} \qquad {\mathsf m}^2_{W'w}\ (w=0,1,\cdots, q-1)~,
\end{equation}
we will show that the lowest eigenvalue of this matrix, ${\mathsf m}^2_{W'0}$,
is light. As before, we will denote this {\it light} eigenvalue by $M^2_{W'}$, and
the distinction between this and the mass matrix will be clear from context.

In terms of the related Case I moose, we may evaluate $[G_{NC}(Q^2)]_{YY}$
directly, and find
\begin{equation}
  [G_{\rm NC}(Q^2)]_{YY}
  = \dfrac{e^2}{Q^2}\dfrac{[Q^2+M_{W'}^2]M_Z^2}{M_{W'}^2[Q^2+M_Z^2]}
    \left[
      \prod_{{w}=1}^{q-1}
      \dfrac{Q^2+{\mathsf m}_{W'{w}}^2}{{\mathsf m}_{W'{w}}^2}
    \right]
     \left[
      \prod_{{z}=1}^{K}
      \dfrac{{\mathsf m}_{Z{z}}^2}{Q^2+{\mathsf m}_{Z{z}}^2}
    \right]
    \left[
      \prod_{{\hat{q}}=q+1}^{K+1}
      \dfrac{Q^2+{\mathsf m}_{{\hat{q}}}^2}{{\mathsf m}_{{\hat{q}}}^2}
    \right].
\label{eq:exact_YY_related}
\end{equation}
Reading off the residues of the poles in Eq.(\ref{eq:exact_YY_related}), we obtain
\begin{equation}
  [\xi_Z]_{YY} 
  = e^2 \dfrac{M_Z^2- M_{W'}^2}{M_{W'}^2}
      \left[\prod_{{w}=1}^{q-1}
      \dfrac{{\mathsf m}_{W'{w}}^2-M_Z^2}{{\mathsf m}_{W'{w}}^2}
      \right] 
      \left[\prod_{{z}=1}^K
      \dfrac{{\mathsf m}_{Z{z}}^2}{{\mathsf m}_{Z{z}}^2-M_Z^2}
      \right]
      \left[\prod_{{\hat{q}}=q+1}^{K+1}
      \dfrac{{\mathsf m}_{{\hat{q}}}^2-M_Z^2}{{\mathsf m}_{{\hat{q}}}^2}
      \right],
\label{eq:Z_residue_YY_related}
 \end{equation}
 
If we expand the expression for $[\xi_Z]_{YY}$ to first non-trivial order, we can rewrite it  as
\begin{equation}
[\xi_Z]_{YY} = \frac{e^2 (M_Z^2 - M_{W'}^2)}{M_{W'}^2} 
[1 + M_Z^2 (\Sigma_Z - \Sigma_{W'} - \Sigma_{q})]~
\label{eq:exp_xi_Z_YY_related}
\end{equation}
where we have defined
\begin{equation}
\Sigma_{W'} \equiv \sum_{w=1}^{q-1} {1\over {\mathsf m}^2_{W'w}}~.
\end{equation}
Equating the two expressions in eqns. (\ref{eq:exp_xi_Z_YY_general}) and
(\ref{eq:exp_xi_Z_YY_related}), we find the relation
\begin{equation}
M^2_{W'} = M^2_W \left[ 1-(M^2_Z-M^2_W)(\Sigma_{W'}-\Sigma_W+\Sigma_M-\Sigma_q)\right]~,
\label{masswprime}
\end{equation}
justifying our assumption that the matrix $M^2_{W'}$ had a light eigenvalue.

\subsection{The Custodial Consistency Conditions and \protect{$\Delta \rho$}}

We now relate the properties of the charged-boson correlation functions
in the original and related Case I moose.  To faciliate the discussion
we define the following general propagators
\begin{equation}
G^{x,y}_{[I,J]}(Q^2)~,\ \ \ G^{x,y}_{[I,J)}(Q^2)~,\ \ \ G^{x,y}_{(I,J]}(Q^2)\ \ \ G^{x,y}_{(I,J)}(Q^2)~,
\label{eq:notationi}
\end{equation}
where, for example,
\begin{equation}
G^{x,y}_{[I,J]}(Q^2) = g_x g_y  \langle x | {1\over Q^2+M^2_{[I,J]}}|y\rangle~,
\label{eq:notationii}
\end{equation}
is the Euclidean space propagator for the correlation function of the
current at site $x$ with that at site $y$ in the linear moose with vector boson mass-squared
matrix $M^2_{[I,J]}$, and similarly for the other correlation functions. Clearly,
we must have that $I \le {x,y} \le J$. 

Using this notation,
$[G_{CC}(Q^2)]_{WW} = G^{p,p}_{[0,N+1)}(Q^2)$ is the
charged-current correlation function in the original moose and 
$[\tilde{G}_{CC}(Q^2)]_{WW} = G^{p,p}_{[0,q)}(Q^2)$ is the charged boson correlation function 
in the related Case I moose. By definition,
\begin{equation}
[G_{CC}(Q^2=0)]_{WW} = g^2_p (M^{-2}_W)_{p,p} \equiv 4\sqrt{2} G_F~,
\label{eq:fac-a}
\end{equation}
while
\begin{equation}
[\tilde{G}_{CC}(Q^2=0)]_{WW} = g^2_p (M^{-2}_{W'})_{p,p} \equiv 4\sqrt{2} \rho G_F~,
\label{eq:fac-b}
\end{equation}
where the second equality follows from
the fact that the zero-momentum charged-current correlation function of the related Case I
moose has the same strength as the $T_3^2$ part of the neutral current correlation function
in the original moose ({\it i.e.} $\rho\equiv 1$ in Case I). 
In these equations $(M^{-2}_{W})_{p,p}$ and $(M^{-2}_{W'})_{p,p}$ represent, respectively, the $(p,p)$ element of the
inverse of the mass-squared matrices $M^2_{W}$ and $M^2_{W'}$. As quoted in Appendix B,
these individual elements are calculated to be
\begin{equation}
  (M_{W}^{-2})_{i,j} = \dfrac{4}{g_i g_j} 
  \sum_{k=\max(i,j)+1}^{N+1} \dfrac{1}{f_k^2}~,
  \qquad
  (M_{W'}^{-2})_{i,j} = \dfrac{4}{g_i g_j} 
  \sum_{k=\max(i,j)+1}^{q} \dfrac{1}{f_k^2}~.
  \label{eq:inverse}
\end{equation}

Then using (\ref{eq:inverse}) in equation (\ref{eq:fac-a}) shows that the weak scale is related to the $f$-constants of the links between the $SU(2)$ group to which the fermions couple ($p$) and the $U(1)$ group at the interface ($N+1$):
\begin{equation}
   \sum^{N+1}_{j=p+1} {4 \over f^2_j}  \equiv {4\over v^2} = 4 \sqrt{2} G_F~,
\end{equation}
while applying (\ref{eq:inverse}) to the difference of equations (\ref{eq:fac-b}) and (\ref{eq:fac-a}) yields a simple expression for $\Delta\rho$:
\begin{equation}
  \Delta\rho = \rho - 1 = \sum_{k=N+2}^{q}
  \dfrac{v^2}{f_k^2} > 0 \ .
  \label{eq:drho_f}
\end{equation}
We see that the low-energy $\rho$ parameter will be identically $1$ 
when $q = N+1$ and will be greater than 1 otherwise.  Furthermore, using these formulae
we find that
\begin{equation}
G^{p,N+1}_{[0,q)}(Q^2=0) = G^{N+1,N+1}_{[0,q)}(Q^2=0) = 4 \sqrt{2} \Delta \rho\, G_F~.
\end{equation}

By direct inversion, we compute the
following correlation functions\footnote{The form of these correlation functions
is determined entirely by the zeros and poles which arise from examining the relevant inverse
matrix, and by fixing the value at $Q^2=0$.} (where the products are over {\it all} non-zero eigenvalues)
\begin{eqnarray}
[G_{CC}(Q^2)]_{WW} & = & {4 \sqrt{2} G_F}{{\prod_{p,r} \left(1+{Q^2\over{\mathsf m}^2_p}\right)
\left(1+{Q^2\over{\mathsf m}^2_r}\right)}\over \prod_w \left(1+{Q^2\over{\mathsf m}^2_{Ww}}{}\right)} 
\label{gwwcc}\\
{}
[\tilde{G}_{CC}(Q^2)]_{WW} & = & {4 \sqrt{2} \rho\, G_F}{{\prod_{p,r'} \left(1+{Q^2\over{\mathsf m}^2_p}\right)
\left(1+{Q^2\over{\mathsf m}^2_{r'}}\right)}\over \prod_{w'} \left(1+{Q^2\over{\mathsf m}^2_{w'}}\right)}~,
\label{gtildewwcc}
\end{eqnarray}
and
\begin{eqnarray}
G^{p,N+1}_{[0,q)}(Q^2) & = & 4\sqrt{2} \Delta \rho\,G_F
{{\prod_{p,s} \left(1+{Q^2\over{\mathsf m}^2_p}\right)
\left(1+{Q^2\over{\mathsf m}^2_s}\right)}\over \prod_{w'} \left(1+{Q^2\over{\mathsf m}^2_{w'}}\right)} \cr
G^{N+1,N+1}_{[0,q)}(Q^2) & = & 4\sqrt{2} \Delta\rho\, G_F
{{\prod_{w,s} \left(1+{Q^2\over{\mathsf m}^2_{Ww}}\right)
\left(1+{Q^2\over{\mathsf m}^2_s}\right)}\over \prod_{w'} \left(1+{Q^2\over{\mathsf m}^2_{w'}}\right)} ~.
\label{glastnc}
\end{eqnarray}
In these expressions, we have defined the matrices and corresponding eigenvalues
\begin{eqnarray}
M^2_{r'} & = & M^2_{(p,q)} \qquad {\mathsf m}^2_{r'}\ (r'=p+1, \cdots, q-1) \\
M^2_{s} & = & M^2_{(N+1,q)} \qquad {\mathsf m}^2_{s}\ (s=N+2,\cdots, q-1)~.
\end{eqnarray}
Phenomenological considerations will  require that these 
eigenvalues be large, ${\mathsf m}^2_{r',s} \gg M^2_{W,Z}$.

The arguments which lead to the consistency condition
of eqn. (\ref{eq:corr-case1}) may be used to show that

\begin{equation}
\tilde{G}^{WW}_{CC}(Q^2) - G^{WW}_{CC}(Q^2)=G^{p,p}_{[0,q)}(Q^2) - G^{p,p}_{[0,N+1)}(Q^2) = 
{[G^{p,N+1}_{[0,q)}(Q^2)]^2 \over G^{N+1,N+1}_{[0,q)}(Q^2)}~.
\label{newconsistency}
\end{equation}
and also to prove the more general consistency conditions listed in Appendix D. Performing a low-$Q^2$ expansion on eqn. (\ref{newconsistency}), we
find the relation
\begin{eqnarray}
4\sqrt{2} \rho\, G_F \left(1+ {Q^2\over M^2_W}\right)(1+Q^2\Sigma_{r'}+Q^2\Sigma_W)
& = & 4\sqrt{2} G_F  \left(1+ {Q^2\over M^2_{W'}}\right)(1+Q^2\Sigma_{r}+Q^2\Sigma_{W'}) \nonumber \\
& + & 4\sqrt{2} \Delta \rho\, G_F (1+Q^2\Sigma_p + Q^2\Sigma_s)~,
\end{eqnarray}
where we have defined
\begin{equation}
\Sigma_{r'} = \sum_{r'=p+1}^{q-1} {1\over {\mathsf m}^2_{r'}}\ , \ \ \ \ \ 
\Sigma_{s} = \sum_{s=N+2}^{q-1} {1\over {\mathsf m}^2_s}\ .
\end{equation}
We know from phenomenologically considerations that $\Delta \rho < {\cal O}(10^{-2})$ --
therefore we may neglect the last terms in this expression, which are proportional to
$\Delta \rho\cdot \Sigma_{p,s} \cdot Q^2$.  Equating powers
of $Q^2$ and $Q^4$ we find (to lowest nontrivial order in deviation from the standard
model):
\begin{eqnarray}
\Sigma_{r'} + \Sigma_W & = & \Sigma_{r} + \Sigma_{W'}~, \label{custodiali} \\
M^2_W & = & \rho\, M^2_{W'}~. \label{custodialii}
\end{eqnarray}

The equations (\ref{custodiali}) and (\ref{custodialii}) are the primary
results of this subsection, and will allow us to compute the electroweak parameters
of the original general model in terms of the properties of the related Case I moose.


\section{General Linear Moose Models:  \protect{$q \geq N+1$}\\ Electroweak Phenomenology}

\subsection{\protect{$\alpha S$} and \protect{$\alpha T$}}

Comparing our pairs of expressions from Sections 4 and 8 for $[\xi_Z]_{YY}$, $[\xi_Z]_{WY}$, $[\xi_Z]_{WW}$, and $[\xi_W]_{WW}$ (after moving consistently to the scheme with $G_F$ as an experimental input) we can solve for $S$ and $T$ to leading order:
\begin{equation}
\alpha S = 4 s_Z^2 c_Z^2 M_Z^2 (\Sigma_Z - \Sigma_p - \Sigma_q)
\label{eq:SS}
\end{equation}
\begin{equation}
\alpha T = s_Z^2 M_Z^2 (\Sigma_Z - \Sigma_W + \Sigma_M - 2 \Sigma_q)~.
\label{eq:TT}
\end{equation}
The linear combination of parameters computed in \cite{Chivukula:2004pk} now looks like
\begin{equation}
\alpha (S - 4 c_Z^2 T) = 4 s_Z^2 c_Z^2 M_Z^2 \left[ (\Sigma_W - \Sigma_p) - (\Sigma_M - \Sigma_q)\right].
\end{equation}
and depends on the differences of the inverse-square-sums of eigenvalues of the full $W$ and $M$ matrices and the reduced-rank $p$ and $q$ matrices.  

If we take $q=N+1$ we recover the equivalent expressions for Case I;  if we take $p=0$ along with $q=N+1$ we recover the expressions from \cite{Chivukula:2004pk}.  
Note that models in which $p=N$ and $q=N+1$ are an extension of the Generalized-BESS-Type-I models studied in \cite{Chivukula:2003wj,Casalbuoni:2004id} and our results are consistent with these earlier papers.  
In this case we have $\Sigma_q = \Sigma_M$, and $\Sigma_p$ takes on a value which we denote $\Sigma_{p=N}$.  Therefore equations (\ref{eq:SS}) and (\ref{eq:TT}) become 
\begin{eqnarray}
\alpha S &=& 4 s_Z^2 c_Z^2 M_Z^2(\Sigma_Z - \Sigma_{p=N} - \Sigma_M)\cr
\alpha T &=& s_Z^2 M_Z^2 (\Sigma_Z -\Sigma_W - \Sigma_M)\cr
\alpha S - 4 c_Z^2 \alpha T &=& 4 s_Z^2 c_Z^2 M_Z^2 (\Sigma_W - \Sigma_{p=N})
\end{eqnarray}
References \cite{Chivukula:2003wj,Casalbuoni:2004id} focused on the case in which $g_p$ and $g_q$ were small.
In this limit, all of the parameters above are of
order ${\cal O}(M^4_W/M^4_H)$, and the first non-zero contributions to electroweak corrections were found to arise as the 4th power of mass ratios rather than the 2nd power.

\subsection{\protect{$\alpha \delta$}}

We find a general expression for $\delta$ by exploiting our ability to compute $\delta$ for any Case I moose, including the Related Case I moose for our general linear moose (shown in figure 3).
In Section 4, we related the non-resonant contribution to four-fermion processes (arising from heavy neutral boson exchange) to the pole residues and masses by

\centerline{\ \hspace{4.5cm}
$\displaystyle \sqrt{2} G_F {\alpha \delta \over s^2 c^2} 
=\sum_{z=1}^{K}\dfrac{[\xi_{Zz}]_{WW}}{{\mathsf m}_{Zz}^2}~. $\hspace{4.5cm} (4.14)}

\noindent{As the matrix $M^2_Z$ is the same in the original moose and its Related Case I moose,
we see that the value of $\delta$ is the same\footnote{In principle,  the value
of $G_F$ in eqn. (4.14) changes to $\rho\, G_F$ in
the Related Case I moose -- however, since $\Delta \rho = {\cal O}(10^{-2})$, this
change is higher order in deviation from the standard model.} in the original moose and 
its Related Case I moose. From our calculation in Section 7 (see (\ref{eq:deltai})), then,
we can write $\delta$ in terms of the $\Sigma_i$ for the Related Case I moose}
\begin{equation}
{\alpha \delta \over c^2}  = -4s^2_Z c^2_Z M^2_Z (\Sigma_{W'}-\Sigma_{r'}-\Sigma_p) 
\end{equation}
and by applying eqn. (\ref{custodiali}) we obtain $\delta$ in terms of the $\Sigma_i$ for the original general linear moose
\begin{equation}
{\alpha \delta \over c^2} = -4s^2_Z c^2_Z M^2_Z (\Sigma_{W}-\Sigma_{r}-\Sigma_p)~,
\label{eq:theexpp}
\end{equation}

\subsection{$\Delta \rho$}

From eqns. (\ref{masswprime}) and (\ref{custodialii}), we find
\begin{equation}
\Delta \rho =  s^2_Z M^2_Z  (\Sigma_{W'}-\Sigma_W+\Sigma_{M}-\Sigma_q)~.
\label{eq:drho_sigma}
\end{equation}
Recall that we also found (equation \ref{eq:drho_f}) that $ \Delta\rho \equiv \rho - 1 =  \displaystyle \sum_{k=N+2}^{q} \dfrac{v^2}{f_k^2} > 0 $.  
The results of Appendix B may be used to show the equivalence of 
eqns. (\ref{eq:drho_f}) and (\ref{eq:drho_sigma}).
Note that for any Case I moose, $\Sigma_W = \Sigma_{W'}$, $\Sigma_{M}
= \Sigma_q$, and $q=N+1$ --- therefore using either expression
we see that $\Delta \rho$ vanishes.

Furthermore, the experimental bounds on $\rho - 1$ place limits on the size of some of the $f_k$.  As shown\footnote{See \protect\cite{Chivukula:2004af} for a discussion of the correspondence 
between the notation of Barbieri et al. and the notation used here.} in \cite{Barbieri:2004qk}, the current limit is $\rho - 1 < .004$.  In a model with $q = N+2$ (i.e., the fermions couple to the $U(1)$ to the right of the one at the $SU(2)$ - $U(1)$ interface), we  find $f_{N+2} > v / \sqrt{.004} \sim $ 3.9 TeV.  In a model with $q > N+2$, each $f_k$ contributing to $\rho - 1$ must be greater than 3.9 TeV.   This suggests that the Case I models, with $\Delta \rho = 0$ will be of greatest phenomenological interest.

\subsection{Zero Momentum Parameters}

Using the results of \cite{Chivukula:2004af} we can write the zero-momentum parameters as 
\begin{eqnarray}
\hat{S} & = & {1\over 4s^2}\left(\alpha S + 4 c^2 (\Delta \rho - \alpha T) + {\alpha \delta \over c^2}\right) \label{eq:def-a}\\
\hat{T} & = & \Delta \rho \\
W & = & {\alpha \delta \over 4 s^2 c^2} \\
Y & = & {c^2 \over s^2} (\Delta \rho - \alpha T)~.
\label{eq:def-d}
\end{eqnarray}
Inserting our expressions for the on-shell parameters in terms of the $\Sigma_i$ from sections 9.1 and 9.2 and applying the
custodial consistency relation of eqn. (\ref{custodiali}), we find
\begin{equation}
\hat{S} = M^2_W \Sigma_{r'} > 0~,
\label{eq:shatgeneral}
\end{equation}
is strictly greater than zero, and 
\begin{eqnarray}
\hat{T} & = & {s^2_Z \over c^2_Z} M^2_W ( \Sigma_{W'}- \Sigma_{W}+ \Sigma_{M}-\Sigma_q) \label{eq:ehs:that} \\
W & = & -M^2_W  (\Sigma_{W'}-\Sigma_{r'}-\Sigma_p) \\
Y & = & -M^2_W(\Sigma_Z - \Sigma_{W'}-\Sigma_q)~.
\end{eqnarray}
The values of $\hat{S}$, $W$ and $Y$ are precisely the
same as the values one would calculate for the Related Case I moose.

To better understand the non-zero value of $\hat{T}$ in the general linear moose (recall $\hat{T} = 0$ in the related Case I moose) we can return to the polarization amplitudes of equations (\ref{bnclagrangian}) and (\ref{bcclagrangian}).
If we consider the related Case I moose shown in fig. \ref{fig:ttwo}, 
the consistency relation of eqn. (\ref{eq:corr-case1}) implies
\begin{equation}
([G_{NC}(Q^2)]_{WW} - [\tilde{G}_{CC}(Q^2)]_{WW})
\cdot [G_{NC}(Q^2)]_{YY}
= ([G_{NC}(Q^2)]_{WY})^2~.
\label{eq:general-consistency}
\end{equation}
From eqn. (\ref{eq:general-consistency}), we compute
\begin{equation}
\Pi_{W^3 W^3}(Q^2) = {-1\over [\tilde{G}_{CC}(Q^2)]_{WW}}~,
\end{equation}
while
\begin{equation}
\Pi_{W^+ W^-}(Q^2) = {-1\over [G_{CC}(Q^2)]_{WW}}~,
\end{equation}
and therefore $\Pi_{W^+ W^-}(Q^2) - \Pi_{W^3 W^3}(Q^2) \neq 0$ so that $\hat{T}$ will not vanish. Using eqn.
(\ref{newconsistency}), and the forms of the correlation functions in eqns. (\ref{gwwcc})
-- (\ref{glastnc}), we compute
\begin{equation}
\Pi_{W^3 W^3}(Q^2) - \Pi_{W^+ W^-}(Q^2) = {\Delta \rho \over 4\sqrt{2} \,\rho\, G_F } 
{ \prod_{s}  \left(1+{Q^2\over{\mathsf m}^2_s}\right) \over
\prod_{r,r'} 
\left(1+{Q^2\over{\mathsf m}^2_r}\right)\left(1+{Q^2\over{\mathsf m}^2_{r'}}\right)}~.
\label{result}
\end{equation}
This expression reproduces the value of $\hat{T}$ in equation (\ref{eq:ehs:that}) when evaluated at $Q^2=0$.  

Given 
that the variables $\hat{U}$ and $V$ depend on higher order derivatives of the
difference in eqn. (\ref{result}),  when the ${\mathsf m}^2_{r,r',s}$ are large
we see that $\hat{U}$ and $V$ will be much smaller than $\hat{T}$.

\section{Which Models are Viable?}

The preceding sections of this paper have analyzed the corrections to precisely measured electroweak
quantities in the moose shown in figure \ref{fig:TheMoose}, subject to the following phenomenologically-motivated assumptions:
\begin{enumerate}

\item The matrix  $M^2_Z = M^2_{[0,N+M+1]}$ has only one light non-zero mass eigenvalue, which is
associated with the ordinary $Z$ boson;
\item The matrices $M^2_{W} = M^2_{[0,N+1)}$ and $M_{W'}^2 = M^2_{[0,q)}$ each have only one light eigenvalue.  The light eigenvalue of $M^2_W$ is associated with the ordinary $W$ boson;
\item None of the submatrices $M^2_p = M^2_{[0,p)}$, $M^2_{r} = M^2_{(p,N+1)}$, $M^2_{r'} = M^2_{(p,q)}$, $M^2_q
= M^2_{(q,K+1]}$, and ${\cal M}^2_{M} = M^2_{(N+1,K+1]}$ has a light
eigenvalue with mass of order $M_W^2$ or $M_Z^2$;
\item The $f$-constants and couplings of the moose are constrained to obey
\begin{equation}
\sqrt{2} G_F = {1\over v^2} = \sum_{k=p+1}^{N+1} \frac{1}{f_k^2}~,
\qquad \qquad {1\over e^2} = \sum_{i=0}^{N+M+1} 
{1\over g^2_i}~.
\label{eq:constraints}
\end{equation}
\end{enumerate}
We find that the deviations in electroweak parameters from their standard model values may be summarized by the following four quantities, each of which is constrained by experiment 
\cite{Barbieri:2004qk} to be less than of order $10^{-3}$ in magnitude:
\begin{eqnarray}
\hat{S} & = & M^2_W \Sigma_{r'} > 0~, \label{eq:shat}\\
\hat{T} & = & {s^2_Z \over c^2_Z} M^2_W ( \Sigma_{W'}- \Sigma_{W}+ \Sigma_{M}-\Sigma_q) \label{eq:that}\\
W & = & -M^2_W  (\Sigma_{W'}-\Sigma_{r'}-\Sigma_p) = -M^2_W  (\Sigma_{W}-\Sigma_{r}-\Sigma_p) \label{eq:W} \\
Y & = & -M^2_W(\Sigma_Z - \Sigma_{W'}-\Sigma_q)~. \label{eq:Y}
\end{eqnarray}
Because the $\Sigma_i$ are sums over the inverse-squares of mass eigenvalues, barring
some enhancement of a subleading contribution that cancels against these terms, 
the above constraints on electroweak parameters imply that the ratio $M_W^2 / {\mathsf m}^2$ for any heavy eigenvalue of any of the mass matrices must likewise be less than or of order $10^{-3}$.

In this section, we explore two further questions.  First, we determine which configurations of the $f_i$ and $g_i$ result in a linear moose satisfying all of the constraints above.  Second, we ask whether such a moose can also be consistent with
unitarity, which places an upper bound on the masses of the extra $W$ bosons that unitarize high-energy longitudinal W-boson scattering.

\subsection{Couplings and $f$-constants}

Let us determine which values of the $f_k$ and $g_i$ will result in a linear moose that meets the phenomenological constraints above (except unitarity which we will address shortly).   As discussed in Appendix B, the mass eigenvalues of the charged-boson portion of the moose and its submatrices are related as
\begin{equation}
\left[\prod_{\hat{p}=1}^{p} {\mathsf m}_{\hat{p}}^2\right] \, g_p^2\, 
\left[\frac{1}{4} F_{W,p}^2 \prod_{{r}=p+1}^{N} {\mathsf m}_{{r}}^2\right]
= M_W^2 \prod_{\hat{w}=1}^{N} {\mathsf m}_{W{\hat{w}}}^2\ ,
\end{equation}
where
\begin{equation}
{1\over F^2_{W,p}} = \sum_{i=p+1}^{N+1} {1\over f^2_i}~.
\end{equation}
Having $g_p$ be small is sufficient to enable the $W$ to be light without having any 
of the ${\mathsf m}_{\hat{p}}^2$ or ${\mathsf m}_{{r}}^2$ be comparably light.  If any other $g_i$ were small, some related ${\mathsf m}_{\hat{i}}^2$ would also be light.   Likewise,  having $g_p$ and $g_q$ small is sufficient to enable the full moose to
have a light $Z$ eigenstate without having any of the ${\mathsf m}_{\hat{p}}^2$, ${\mathsf m}_{\hat{q}}^2$, or ${\mathsf m}_{{r}}^2$ be similarly light.    This demonstrates the existence of acceptable configurations of the $f_k$ and $g_i$.  We would like to go further and determine the nature of all acceptable configurations.  We will focus on the charged-boson portion of the linear moose (fig. \ref{fig:ffour}) because it will directly relate to our eventual concerns about satisfying unitarity bounds.
\EPSFIGURE[ht]{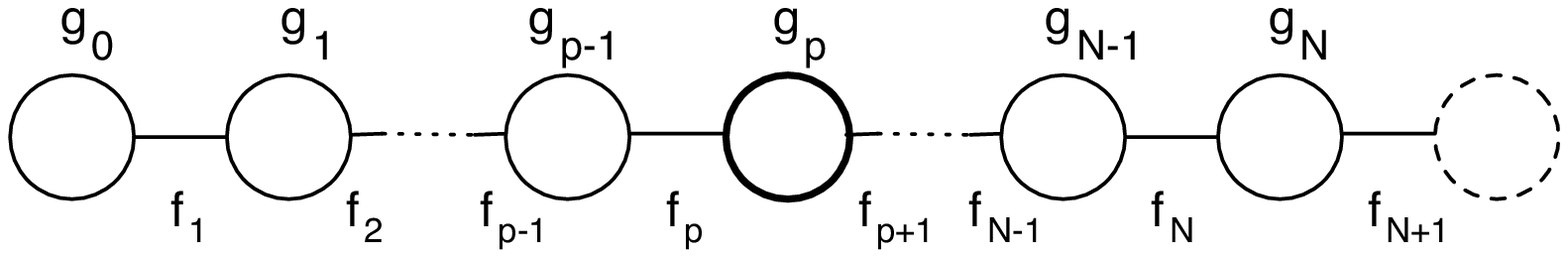,width=0.8\textwidth}
{Moose diagram corresponding to the charged-boson mass
matrix $M^2_W = M^2_{[0,N+1)}$.
\label{fig:ffour}}

 Our discussion employs the intuition derived from Georgi's
``spring analogy'' \cite{Georgi:2004iy}, which is reviewed in Appendix E. 
Namely, we know that the eigenvalues of
the mass-squared matrix for any linear moose correspond to those of a system of coupled
springs with masses $1/g^2_i$ and spring constants $f^2_k$.  The amplitudes
of the mass-squared eigenvectors correspond to  the normal-mode displacements
 of the spring system. Note that we must satisfy
the constraints of eqn. (\ref{eq:constraints}), and therefore the couplings obey $g_i >e$ (for
all $i$) and the $f$-constants obey $f_k > v$ ($k=p+1, \cdots, N+1$). The global symmetry at
site $N+1$ is a fixed boundary, equivalent to an infinite mass or $g^2_{N+1} \equiv 0$.
Satisfying the condition that $M^2_W/{\mathsf m}^2< {\cal O}(10^{-3})$ corresponds to our
system's having one and only one low-frequency mode.   We proceed by examining several possibilities  in turn.

If all of the couplings and $f$-constants are comparable, with no hierarchies among them, there
is no way to establish a large hierarchy between the lightest eigenvalue and the next. This is 
analogous to the KK spectrum for the gauge fields in a flat background with spatially independent
coupling: the tower of KK states is linear in mass-squared.   This is not phenomenologically acceptable.

Suppose, instead, that one of the $g_i$ is smaller than all the rest, but all the $f_k$'s are comparable 
in size. Such a situation clearly has one light eigenvalue -- 
in the spring analogy, there is a low-frequency mode in which site $i$ moves slowly and there are approximate 
heavy modes with site $i$ remaining still.   Then there will be a single light eigenstate of $M_W^2$.  However,
recall that  all of the parameters in eqns. (\ref{eq:shat}-\ref{eq:Y}) are constrained to be small. Examining
eqn. (\ref{eq:W}), we see that neither of the matrices ${\cal M}^2_p=M^2_{[0,p)}$ and ${\cal M}^2_r= M^2_{(p,N+1)}$ may 
have a mass eigenvalue of order $M_W^2$, or else the parameter $W$ would be larger than experiment allows (since $\Sigma_W$ does not include the light $W$ mass and could not cancel a large $\Sigma_r$ or $\Sigma_p$).
If the small coupling $g_i$ corresponds to a site with $i<p$, the matrix ${\cal M}^2_p$ will have a small
eigenvalue; if $i>p$, the matrix $M^2_r$ will have a small eigenvalue. We therefore
conclude that $g_i = g_p$, is the only viable option when all of the $f_k$ are of similar magnitude.  

A small extension of the above analysis shows that we cannot have hierarchical $g_i$'s
with more than one small coupling when all $f_k$ are comparable. Suppose there are 
two couplings much smaller than the
rest: $g_{i,i'}$. The high-frequency, large mass-squared, modes are those where the displacements
at sites $i,i'$ are approximately zero.  Conversely, there are now two low-frequency modes involving
the displacements at sites $i,i'$, implying there is more than one light mass-squared
eigenvalue for matrix $M_W^2$.  This contradicts observation.  

Now let us look at the opposite extreme in which all of the $g_i$ are comparable and one
of the $f_k$ is small.  Since the $f$-constants link pairs of gauge groups, it is not possible for an
$f_k$ to form part of matrix $M_W^2$ without being included in either ${\cal M}_p^2$ or ${\cal M}_r^2$.  Hence,
an $f_k$ small enough to produce a light eigenvalue in $M_W^2$ will also produce a small
eigenvalue in one of the submatrices.

The related case in which all of the $g_i$ are comparable and two or more $f_k$ are small
is also forbidden.  To see this, consider the dual moose \cite{Sfetsos:2001qb} of our charged-gauge-boson moose in which the roles of sites ($g_i$) and links ($f_i$) are exchanged.
As reviewed in Appendix E,  this moose has the same mass-squared 
eigenvalues as the original moose.  In the dual moose, we would have all $f_i$ comparable and two or more $g_k$ small.  But we saw earlier that
this necessarily leads to more than one low-frequency eigenmode.

The only viable case identified so far is one in which all of the $f_k$ are of comparable size
and the coupling $g_{i=p}$ is small.  We now ask whether introducing hierarchies in the values of
the $f_k$ would allow one to have the small coupling at a site $g_{i\neq p}$.  Again, the goal is
to have a mass-spring system with one heavy mass whose motion provides a single low-frequency
eigenmode in $M_W^2$ and no such eigenmodes in ${\cal M}_r^2$ or ${\cal M}_p^2$.   Suppose that the
small coupling is at site $i\neq p$ and that all of the $f$-constants linking sites between $i$ and $p$ are large; this is equivalent to having a heavy mass at $i$ connected to the mass at site $p$ by very stiff springs.  The masses between $i$ and $p$ will oscillate as a single heavy unit; if one integrated out the large $f$-constants, one would obtain an effective site with a smaller coupling
than the original $g_i$.  
Because the fermions originally coupled to site $p$, which is subsumed in the new effective low-coupling site $p^{eff}$, the new effective site will play the role of the original site $p$.  In particular, coupling $g_{p,eff}$ appears in the $W$ mass-squared matrix, but not in ${\cal M}^2_{p,eff}$ or ${\cal M}^2_{r,eff}$.  Hence, the large $f$-constants have made a situation with small $g_{i\neq p}$ viable.

Note that several extensions are possible.  The chain of large $f_k$ must include both site $i$ and site $p$, but neither
of those sites is required to be an endpoint.  Also, it is possible to have more than one small-coupling site among those
linked to $i$ and $p$ by large $f$-constants.  The constraint  (\ref{eq:constraints}) on the values of the $f_k$ lying between sites $p$ and $N+1$ can always be satisfied by making link $f_{N+1}$ of order $v$ since site $N+1$ is associated with a $U(1)$ group and does not contribute to $M_W^2$ or its submatrices. Finally, it is possible to have one or more small-coupling sites embedded in a
chain of large $f$-constants which includes the right-most link $f_{N+1}$ -- this corresponds
in the oscillator language to a ``thick'' wall for the fixed boundary.

\subsection{Unitarity}

Having identified which configurations of $f_k$ and $g_i$ can in principle satisfy the precision electroweak constraints, we now investigate whether they can simultaneously be compatible with unitarity bounds.  

Recall that $\sqrt{8\pi} v$ is the scale at which $W_L W_L$ spin-0 isospin-0 scattering would
violate unitarity in the absence of a Higgs boson. From the equivalence theorem, we know that the
scattering of high-energy longitudinal $W$ bosons
is identical to the scattering of the $R_\xi$-gauge pions eaten by the $W$-bosons. The eaten pions are necessarily a linear
combination of the pions implicit in the $N+1$ $SU(2)\times SU(2)/SU(2)$ nonlinear sigma-model
``links'' in the moose shown in figure 1. Unitarity, therefore, guarantees that the
mass-squared $M^2_{W1}$, of the next lightest state after the $W$-boson, is bounded
from above by $8\pi v^2$.  

\EPSFIGURE[ht]{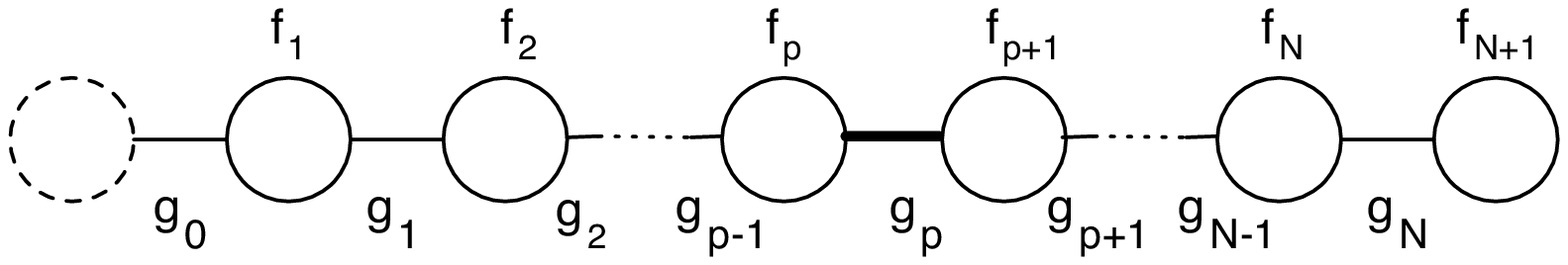,width=0.8\textwidth}
{Dual Moose diagram corresponding to Goldstone bosons eaten
by the charged vector-bosons.  The bold link labeled \protect{$g_p$} corresponds
to the site to which fermions couple in the original moose.
\label{fig:tthree}}

In the phenomenologically relevant linear mooses defined in the previous subsection, the pion eaten by
the $W$-boson is largely a combination of the pions coming from links $p+1$ to $N+1$. 
This is easiest to see in terms of the dual moose. As shown in Appendix E,
the mass-squared matrix of the dual moose
corresponds to the $R_\xi$-gauge mass-matrix of the eaten Goldstone bosons  -- precisely the
objects related to high-energy longitudinal vector-boson scattering via the equivalence theorem.

Consider first the case of all $f$-constants comparable and one small coupling, $g_p \ll g_{i\neq p}$.
The dual moose, as shown in Fig. \ref{fig:tthree}, corresponds to a spring-mass system with 
a fixed boundary at the left-hand side and a ``weak'' spring corresponding to the
small $g_p$. The smallest
eigenvalue, which corresponds to the Goldstone boson eaten by the $W$, arises from
a normal mode in which the masses corresponding to the sites from $f_{p+1}$ to
$f_{N+1}$ move in unison while those at sites $f_1$ to $f_p$ do not -- and the only spring being
stretched corresponds to the coupling $g_p$. The amplitudes of the displacements of this
normal mode correspond to the Goldstone boson eaten by the $W$, which is therefore made up entirely (to leading
order) of the pions coming from links $p+1$ to $N+1$. 

In the limit $g_p \ll g_{i\neq p}$, the moose in Fig. \ref{fig:ffour} approximately cleaves at site $p$ into
separate mooses with mass-squared matrices ${\cal M}_p^2$ and ${\cal M}_r^2$.  Hence, in this
limit, the heavy mass eigenstates of $M_W^2$ are approximately given by the 
eigenstates of ${\cal M}^2_p$ and ${\cal M}^2_r$. As the pions corresponding to the longitudinal
polarization states of the $W$ come almost entirely from sites $p+1$ to $N+1$, their
scattering can only be unitarized by the vector bosons coming from ${\cal M}^2_r$.
Hence, the unitarity bound also applies to the lightest eigenstate of ${\cal M}^2_r$: 
\begin{equation}
\Sigma_r > {1\over 8 \pi v^2}~.
\label{eq:bound}
\end{equation}

The case in which $g_0$ is small is discussed in 
detail in Appendix F; as anticipated by the argument of the previous subsection, it is 
shown explicitly that satisfying the precision electroweak constraints (specifically, keeping $W$ or $\delta$ small) requires that the $f_k$ between sites $0$ and $p$ be large. Therefore, the
pion eaten by the light $W$ is concentrated on links $f_{p+1}$ through $f_{N+1}$.  
Again, the scattering of such a pion  can only be unitarized by a vector meson 
concentrated at sites $p+1$ through $N$.  And, again, because the masses from 
sites $0$ to $p$ remain approximately fixed in the high frequency
modes, the second lightest eigenstate of $M_W^2$ is approximately an eigenstate of ${\cal M}^2_r$
and the bound of eqn. (\ref{eq:bound}) applies.  

This analysis applies directly to any of the realistic cases considered in the
previous subsection, as any weak-coupling sites must be embedded in a chain
of adjacent links with large $f$-constants which includes either  site $p$ or
link $N+1$.  Unitarity therefore
tells us that $M_W^2\Sigma_r \geq 0.004$. The sum $\Sigma_{r'}$ appearing 
in parameter $\hat{S}$  (eqn. (\ref{eq:shat})) is strictly larger than
$\Sigma_r$, as can be shown by direct evaluation of these sums using the explicit form of the inverse matrices (in Appendix B). We therefore conclude that unitarity requires 
$\hat{S} \geq 0.004$, a value too large to be phenomenologically acceptable. 
Moose models  which do not have extra light vector bosons (with masses of the order of the 
$W$ and $Z$ masses) cannot simultaneously satisfy the constraints of precision electroweak data and unitarity bounds.

\section{Discussion and Summary}

In this paper, we have  calculated the form of the corrections to 
the electroweak interactions in the class of Higgsless models which can be 
``deconstructed'' to a  chain of $SU(2)$ gauge groups adjacent to a chain of $U(1)$ gauge groups, 
and with the fermions coupled to any single $SU(2)$ group and to any single 
$U(1)$ group along the chain. We have related the size of corrections to electroweak processes
in these models to the spectrum of vector bosons which, in turn,
is constrained by unitarity in Higgsless models. 
In particular, we find that the electroweak parameter
$\hat{S}$ is bounded by
\begin{equation}
\hat{S} ={1\over 4s^2}\left(\alpha S + 4 c^2 (\Delta \rho - \alpha T) + {\alpha \delta \over c^2}\right) 
\ge M^2_W \Sigma_r \ge {M^2_W\over 8 \pi v^2} \simeq 4 \times 10^{-3}~,
\label{eq:final}
\end{equation}
which is disfavored by  precision electroweak data. We have shown that
this bound is true for arbitrary background 5-D geometry, 
spatially dependent gauge-couplings, and brane kinetic energy terms. 

Although we have stressed our results as they apply to continuum Higgsless
5-D models, they apply to any linear moose model including models motivated by
hidden local symmetry. Our calculations also apply directly to the electroweak 
gauge sector of 5-D theories with a bulk
scalar Higgs boson, although the constraints arising from unitarity (the
right-most inequality in eqn. (\ref{eq:final}) above) no longer applies.
The extension of these calculations to consider delocalized fermions
is under investigation.

 \acknowledgments

We would like to thank Nick Evans, Howard Georgi, and
John Terning for discussions. R.S.C. and E.H.S. gratefully acknowledge the
hospitality of the Aspen Center for Physics, where some of this work was completed.
M.K. acknowledges support by the 21st Century COE Program of Nagoya University 
provided by JSPS (15COEG01). M.T.'s work is supported in part by the JSPS Grant-in-Aid for Scientific Research No.16540226. H.J.H. is supported by the US Department of Energy grant
DE-FG03-93ER40757.


\appendix

\section{Appendix:  The Sub-matrices \protect{${\cal M}^2_{p,q,r,L}$}}

For the reader's convenience, we write out explicitly the matrices ${\cal M}_{p,q,r,L}$ introduced in Section 3:

\begin{equation}
{\tiny
4 {\cal{M}}_{p}^2 = 4 {{M}}_{[0,p)}^2 = 
\left(
\begin{array}{c|c|c|c|c|c}
g^2_0 f^2_1& -g_0 g_1 f^2_1 & & &  \\ \hline
-g_0 g_1 f^2_1  & g^2_1(f^2_1+f^2_2) & -g_1 g_2 f^2_2 & &   \\ \hline
 & -g_1 g_2 f_2^2 & g^2_2(f^2_2+f^2_3) & -g_2 g_3 f^2_3 &  \\ \hline
 & & \ddots & \ddots & \ddots &   \\ \hline
  & & & -g^{}_{p-3} g^{}_{p-2} f^2_{p-2} & g^2_{p-2}(f^2_{p-2}+f^2_{p-1}) & -g^{}_{p-2} g^{}_{p-1} f^2_{p-1} \\ \hline
 & & & & -g^{}_{p-2} g^{}_{p-1} f^2_{p-1} & g^2_{p-1}(f^2_{p-1}+f^2_{p})\\
\end{array}
\right)
}
\label{chargedmatrix2}
\end{equation}
\begin{eqnarray}
\lefteqn{
4{\cal M}_{r}^2 = 4 {M}_{(p,N+1)}^2 = }\nonumber \\
& & {\tiny \left(
\begin{array}{c|c|c|c|c|c}
g_{p+1}^2 (f_{p+1}^2+f_{p+2}^2)   & -g^{}_{p+1} g^{}_{p+2} f_{p+2}^2      &
   &  &  & \\ \hline
-g^{}_{p+1} g^{}_{p+2} f_{p+2}^2        & g_{p+2}^2 (f_{p+2}^2+f_{p+3}^2) & 
-g^{}_{p+2} g^{}_{p+3} f_{p+3}^2  &  &  & \\ \hline
                                  & -g^{}_{p+2} g^{}_{p+3} f_{p+3}^2      &
 g_{p+3}^2 (f_{p+3}^2+f_{p+4}^2)  & -g^{}_{p+3} g^{}_{p+4} f_{p+4}^2      &
                                  & \\ \hline
 & & \ddots & \ddots & \ddots  &   \\ \hline
 & & & -g^{}_{N-2} g^{}_{N-1} f_{N-1}^2         &
          g_{N-1}^2(f_{N-1}^2+f_{N}^2) &  -g^{}_{N-1} g^{}_{N} f_{N}^2       \\ \hline
 & & & & -g^{}_{N-1} g^{}_{N} f_{N}^2  &   g_{N}^2 (f^2_{N}+ f^2_{N+1})         \\
\end{array}
 \right)
 } 
 \nonumber\\
 & & 
 \label{middlematrix}
 \end{eqnarray}
\begin{eqnarray}
  \lefteqn{
      4{\cal M}_q^2 = 4 {M}_{(q,K+1]}^2 = 
  } \nonumber\\
 & & {\tiny
\left(
\begin{array}{c|c|c|c|c|c}
g_{q+1}^2 (f_{q+1}^2+f_{q+2}^2)   & -g^{}_{q+1} g^{}_{q+2} f_{q+2}^2      &
   &  &  & \\ \hline
-g^{}_{q+1} g^{}_{q+2} f_{q+2}^2        & g_{q+2}^2 (f_{q+2}^2+f_{q+3}^2) & 
-g^{}_{q+2} g^{}_{q+3} f_{q+3}^2  &  &  & \\ \hline
                                  & -g^{}_{q+2} g^{}_{q+3} f_{q+3}^2      &
 g_{q+3}^2 (f_{q+3}^2+f_{q+4}^2)  & -g^{}_{q+3} g^{}_{q+4} f_{q+4}^2      &
                                  & \\ \hline
 & & \ddots & \ddots & \ddots  &   \\ \hline
 & & & -g^{}_{K-1} g^{}_{K} f_K^2         &
          g_K^2(f_K^2+f_{K+1}^2) &  -g^{}_K g^{}_{K+1} f_{K+1}^2       \\ \hline
 & & & & -g^{}_K g^{}_{K+1} f_{K+1}^2  &   g_{K+1}^2 f^2_{K+1}         \\
\end{array}
\right)
}
\nonumber\\
 & &
\label{eq:mass_matrixMq}
\end{eqnarray}
\begin{eqnarray}
\lefteqn{
4{{\cal M}}_{L}^2 = 4 {M}_{(p,K+1]}^2 = } \nonumber \\
& & {\tiny
\left(
\begin{array}{c|c|c|c|c|c}
g_{p+1}^2 (f_{p+1}^2+f_{p+2}^2)   & -g^{}_{p+1} g^{}_{p+2} f_{p+2}^2      &
   &  &  & \\ \hline
-g^{}_{p+1} g^{}_{p+2} f_{p+2}^2        & g_{p+2}^2 (f_{p+2}^2+f_{p+3}^2) & 
-g^{}_{p+2} g^{}_{p+3} f_{p+3}^2  &  &  & \\ \hline
                                  & -g^{}_{p+2} g^{}_{p+3} f_{p+3}^2      &
 g_{p+3}^2 (f_{p+3}^2+f_{p+4}^2)  & -g^{}_{p+3} g^{}_{p+4} f_{p+4}^2      &
                                  & \\ \hline
 & & \ddots & \ddots & \ddots  &   \\ \hline
 & & & -g^{}_{K-1} g^{}_{K} f_{K}^2         &
          g_{K}^2(f_{K}^2+f_{K+1}^2) &  -g^{}_{K} g^{}_{K+1} f_{K+1}^2       \\ \hline
 & & & & -g^{}_{K} g^{}_{K+1} f_{K+1}^2  &   g_{K+1}^2 f^2_{K+1}         \\
\end{array}
 \right)
 }
 \nonumber \\
 & &
 \label{Lmatrix}
 \end{eqnarray}
%


\section{Appendix: Mass-Squared Matrix Inverse \& Other Identities}

This appendix contains relationships between the couplings and $f$-constants of linear moose models with either gauged or global groups at their endpoints and the mass eigenvalues of these same mooses.  The relationships are written as for mooses with $N+2$ groups, but can clearly
be adapted for application to the sub-mooses of varying length into which our analysis divides our general linear moose model.

\DOUBLEFIGURE[t]{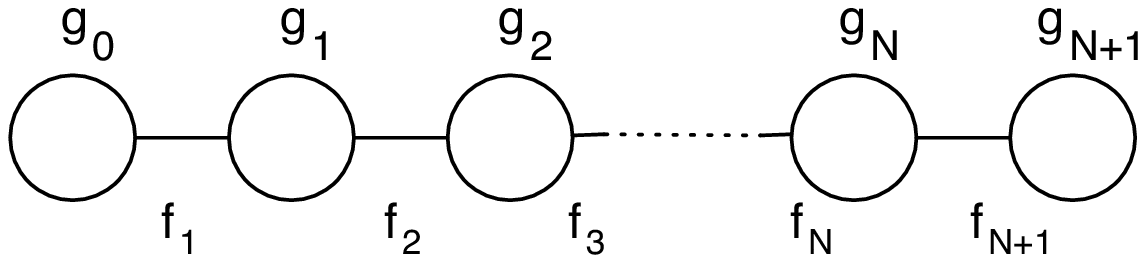,width=0.4\textwidth}
{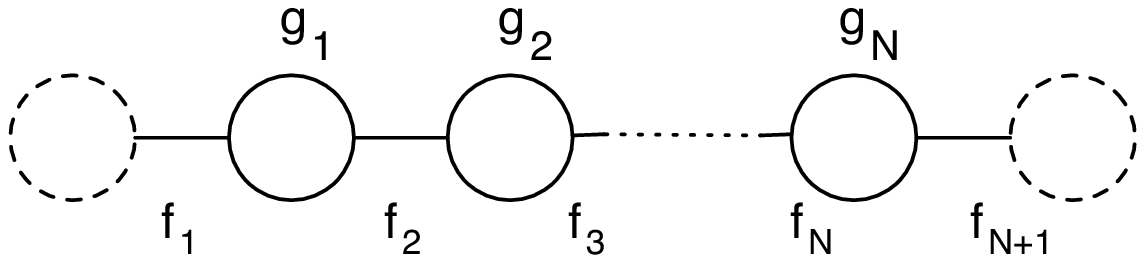,width=0.4\textwidth}
{Linear moose with vector boson mass-squared matrix $M^2_{[0,N+1]}$.
\label{fig:closedmoose}}
{Linear moose with vector boson mass-squared matrix $M^2_{(0,N+1)}$.
\label{fig:openmoose}}

\DOUBLEFIGURE[t]{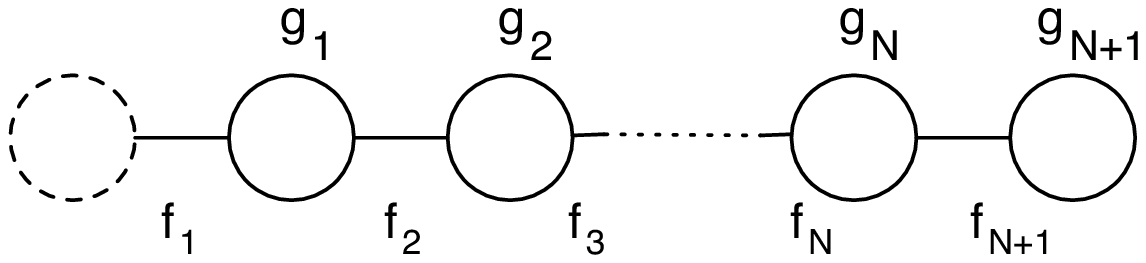,width=0.4\textwidth}
{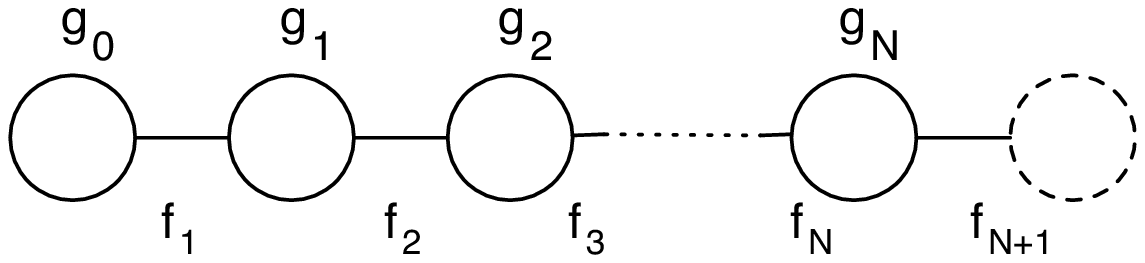,width=0.4\textwidth}
{Linear moose with vector boson mass-squared matrix $M^2_{(0,N+1]}$.
\label{fig:leftopenmoose}}
{Linear moose with vector boson mass-squared matrix $M^2_{[0,N+1)}$.
\label{fig:rightopenmoose}}

Consider the vector-boson mass-squared matrices for the moose models (each with N+2 groups) shown in Figs. (\ref{fig:closedmoose}) - (\ref{fig:rightopenmoose}). Defining
\begin{eqnarray}
{1\over F^2} & = & \sum_{l=1}^{N+1} {1\over f^2_l}~, \\
{1\over F^2_i} & = & \sum_{l=i+1}^{N+1} {1\over f^2_l}~, \\
{1\over \tilde{F}^2_i} & = & \sum_{l=1}^{i} {1\over f^2_l}~,
\end{eqnarray}
the elements of the corresponding nonsingular inverse mass-squared matrices may be written
\begin{eqnarray}
\left\{ M^{-2}_{[0,N+1)} \right\}_{i,j} & = & {4 \over g_i g_j F^2_{{\rm max}(i,j)}} ~, \\
\left\{ M^{-2}_{(0,N+1]} \right\}_{i,j} & = & {4 \over g_i g_j \tilde{F}^2_{{\rm min}(i,j)}} ~, \\
\left\{ M^{-2}_{(0,N+1)} \right\}_{i,j} & = & {4 F^2 \over g_i g_j F^2_{{\rm max}(i,j)} \tilde{F}^2_{{\rm min}(i,j)}}~.
\end{eqnarray}

Denoting the eigenvalues of the matrix $M^2_{[0,N+1)}$ by $[{\mathsf m}^2_{[0,N+1)}]^\ell$, 
and similarly for the other matrices, we have the relations
\begin{eqnarray}
\prod_{\ell=0}^{N} [{\mathsf m}^2_{[0,N+1)}]^\ell & = & 
\prod_{m=0}^{N} g^2_m \, \prod_{p=1}^{N+1} {f^2_p \over 4}~, \label{beq:a}\\
\prod_{\ell=1}^{N+1} [{\mathsf m}^2_{(0,N+1]}]^\ell & = & 
\prod_{m=1}^{N+1} g^2_m \, \prod_{p=1}^{N+1} {f^2_p \over 4}~, \label{beq:b}\\
\prod_{\ell=1}^{N} [{\mathsf m}^2_{(0,N+1)}]^\ell & = & 
{4 \over F^2}\, \prod_{m=1}^{N} g^2_m \, \prod_{p=1}^{N+1} {f^2_p \over 4}~, \label{beq:c}\\
\prod_{\ell \neq 0} [{\mathsf m}^2_{[0,N+1]}]^\ell & = & 
{1\over g^2} \, \prod_{m=0}^{N+1} g^2_m \, \prod_{p=1}^{N+1} {f^2_p \over 4}~, \label{beq:d}
\end{eqnarray}
where the last product is understood to run only over the non-zero eigenvalues
of $M^2_{[0,N+1]}$, and where we have defined
\begin{equation}
{1\over g^2} = \sum_{m=0}^{N+1} {1\over g^2_m}~.
\end{equation}
It is interesting to note that the forms of equations (\ref{beq:c}) and (\ref{beq:d}) are dual to one another (in the sense of exchanging couplings for $f$-constants \cite{Sfetsos:2001qb}, as discussed in the next Appendix E), while the equations in (\ref{beq:a}) and (\ref{beq:b}) are each self-dual in this sense.

One can apply these relations to illustrate that it is possible to ensure that the 
eigenstates of the full linear moose contain a light $Z$ and $W$ (and a massless photon) while the ${\mathsf m}_{Z\hat{k}}$, ${\mathsf m}_{W\hat{w}}$, ${\mathsf m}_{\hat{p}}$, ${\mathsf m}_{{r}}$, and ${\mathsf m}_{\hat{q}}$ are all significantly more massive.   Consider what happens to our general linear moose when $g_p, g_q \equiv 0$.  As shown in figure \ref{fig:ThreeMoose},
groups $p$ and $q$ become global, dividing the moose into three separate mooses of kinds represented in Figures 7-9.  Then we can rewrite the relationship between the couplings, $f$-constants and mass eigenvalues of the full moose in terms of mass eigenvalues of the sub-mooses:
\begin{equation}
e^2 M_Z^2 \prod_{\hat{z}=1}^{K+1} {\mathsf m}_{Z{\hat{z}}}^2 = 
\prod_{\hat{i}=0}^{K+1} g_{\hat{i}}^2 \prod_{\hat{j}=1}^{K+1} \frac{1}{4} f_{\hat{j}}^2 = 
\left[\prod_{\hat{p}=1}^{p} {\mathsf m}_{\hat{p}}^2\right] \, g_p^2\, 
\left[\prod_{\hat{q}=q+1}^{K+1} {\mathsf m}_{\hat{q}}^2\right]\,  g_q^2 \, 
\left[\frac{1}{4} \bar{F}^2 \prod_{{r'}=p+1}^{q-1} {\mathsf m}_{{r'}}^2 \right]~,
\label{eq:ehs-fullmoose}
\end{equation}
where
\begin{equation}
{1\over \bar{F}^2} = \sum^q_{i=p+1} {1\over f^2_i}~.
\end{equation}
Clearly, having $g_p$ and $g_q$ small is sufficient to enable the full moose to
have a light $Z$ eigenstate without having any of the ${\mathsf m}_{\hat{p}}^2$, ${\mathsf m}_{\hat{q}}^2$, or ${\mathsf m}_{{r}}^2$ be light.   One can similarly show that  small $g_p$ is sufficient to enable the $W$ to be light without having any 
of the ${\mathsf m}_{\hat{p}}^2$ or ${\mathsf m}_{{r'}}^2$ be light.  

\EPSFIGURE[t]{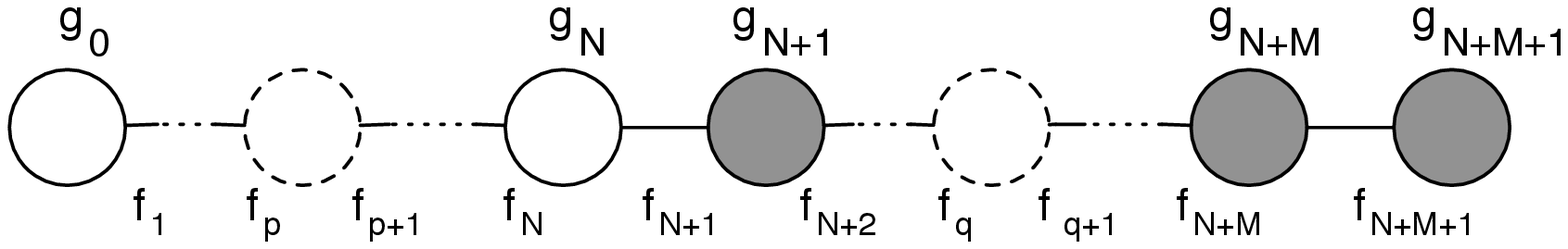,width=0.9\textwidth}
{General moose in limit $g_{p,q} \equiv 0$. The dashed
circles represent the global symmetry groups remaining when $g_{p,q} \to 0$,
and are to be understood to (each) belong to both the smaller moose to the immediate left
and to the immediate right.
\label{fig:ThreeMoose}}

\section{Appendix: Heavy $Z$ and $W$ Boson Pole Residues}

We collect here, for completeness, the residues of the poles corresponding
to the $Z'$ and $W'$ bosons in the correlation functions derived in Sections 5, 6, and 8.

\subsection{Case I Moose}

Correlation function $[G_{NC}(Q^2)]_{WY}$ as written in eqn. (\ref{gncwy}) has the following heavy pole residue in Case I
\begin{equation}
  [\xi_{Z{k}}]_{WY} =
    e^2 \dfrac{M_Z^2}{{\mathsf m}_{Z{k}}^2- M_Z^2}
    \left[\prod_{{z}\ne {k}}
         \dfrac{{\mathsf m}_{Z{z}}^2}
               {{\mathsf m}_{Z{z}}^2  - {\mathsf m}_{Z{k}}^2 }
    \right]
    \left[\prod_{\hat{p}=1}^{p} 
         \dfrac{{\mathsf m}_{\hat{p}}^2 - {\mathsf m}_{Z{k}}^2}
               {{\mathsf m}_{\hat{p}}^2}
    \right]
    \left[\prod_{m=1}^{M} 
         \dfrac{{\mathsf m}_{m}^2 - {\mathsf m}_{Z{k}}^2}
               {{\mathsf m}_{m}^2}
    \right].
\end{equation}

\noindent Correlation function $[G_{NC}(Q^2)]_{YY}$ as written in eqn.(\ref{eq:exact_YY}) has the following heavy pole residue in Case I
\begin{equation}
  [\xi_{Z{k}}]_{YY} =
   - e^2 \dfrac{[{\mathsf m}_{Z{k}}^2 - M_W^2]M_Z^2}
             {M_W^2 [{\mathsf m}_{Z{k}}^2 - M_Z^2]}
      \left[\prod_{{w}=1}^N
      \dfrac{{\mathsf m}_{W{w}}^2-{\mathsf m}_{Z{k}}^2}{{\mathsf m}_{W{w}}^2}
      \right]
      \left[\prod_{{z} \ne {k}}
      \dfrac{{\mathsf m}_{Z{z}}^2}{{\mathsf m}_{Z{z}}^2-{\mathsf m}_{Z{k}}^2}
      \right]
      \left[\prod_{{m}=1}^M
      \dfrac{{\mathsf m}_{{m}}^2-{\mathsf m}_{Z{k}}^2}{{\mathsf m}_{{m}}^2}
      \right]. \nonumber \\
\label{eq:Z_ell_residue_YY}
\end{equation}

\noindent Correlation functions $[G_{CC}(Q^2)]_{WW}$ and $[G_{NC}(Q^2)]_{WW}$ as written, respectively, in eqns. (\ref{eq:exact_gcc_ww})  and (\ref{eq:exact_gnc_ww}) have the following heavy pole residues in Case I
\begin{eqnarray}
[\xi_{Wk}]_{WW} & = & {4\sqrt{2} G_F M^2_W {\mathsf m}^2_{Wk}\over [M^2_W - {\mathsf m}^2_{Wk}]}
\left[ \prod_{w\neq k} {{\mathsf m}^2_{Ww} \over  {\mathsf m}^2_{Ww} - {\mathsf m}^2_{Wk} }\right]
\left[ \prod_{\hat{p}=1}^p {{\mathsf m}^2_{\hat{p}} -{\mathsf m}^2_{Wk} \over  {\mathsf m}^2_{\hat{p}}}\right] \nonumber \\
& \times & 
\left[ \prod_{r=p+1}^{N} {{\mathsf m}^2_{r} -{\mathsf m}^2_{Wk}  \over  {\mathsf m}^2_{r}}\right]~,
\label{eq:Wh_residue_WW_new} \\
{}
[\xi_{Zk}]_{WW} & = & -e^2\,{M^2_Z [M^2_L - {\mathsf m}^2_{Zk}]\over 
M^2_L[M^2_Z-{\mathsf m}^2_{Zk}]}
\left[ \prod_{z\neq k} {{\mathsf m}^2_{Zz} \over  {\mathsf m}^2_{Zz} - {\mathsf m}^2_{Zk}}\right]
\left[ \prod_{\hat{p}=1}^p {{\mathsf m}^2_{\hat{p}} - {\mathsf m}^2_{Zk} \over  {\mathsf m}^2_{\hat{p}}}
\right] \nonumber \\
& \times & 
\left[ \prod_{l=1}^{K-p} {{{\mathsf m}^2_{l} -{\mathsf m}^2_{Zk} }\over  {\mathsf m}^2_l}
\right]~.
\label{eq:Zh_residue_WW_new}
\end{eqnarray}

\noindent The difference of correlation functions $[G_{NC}(Q^2)]_{WW} - [G_{CC}(Q^2)]_{WW}$ as written in eqn. (\ref{eq:exact_WWff}) has the following heavy pole residues in Case I
\begin{eqnarray}
  {[\xi_{W{k}}^{~}]}_{WW}^{~} &=&
  \dfrac{e^2 M_W^2M_Z^2}{(M_W^2 - {\mathsf m}_{W{k}}^2)(M_Z^2 - {\mathsf m}_{W{k}}^2)\,}
   \left[\prod_{{w} \neq {k}}^{N}
     \dfrac{{\mathsf m}_{W{w}}^2}
           {\,{\mathsf m}_{W{w}}^2 \!-\! {\mathsf m}_{W{k}}^2\,}\right]
   \left[\prod_{{z} =1}^{K} 
     \dfrac{{\mathsf m}_{Z{z}}^2}
           {\,{\mathsf m}_{Z{z}}^2 \!-\! {\mathsf m}_{W{k}}^2\,}\right]
\nonumber \\
& \times&
   \left[\prod_{{m} =1}^{M}
     \dfrac{\,{\mathsf m}_{{m}}^2 - {\mathsf m}_{W{k}}^2\,}{{\mathsf m}_{{m}}^2}\right]
   \left[\prod_{\hat{p} =1}^{p} 
     \dfrac{\,({\mathsf m}_{\hat{p}}^2 - {\mathsf m}_{W{k}}^2)^2\,}
           {{\mathsf m}_{\hat{p}}^4}\right] \,,
\label{eq:Wh_residue_WW}
\end{eqnarray}
\begin{eqnarray}
  {[\xi_{Z{k}}^{~}]}_{WW}^{~}  &=&
  -\dfrac{e^2 M_W^2M_Z^2}{\,(M_W^2 -{\mathsf m}_{Z{k}}^2)(M_Z^2 -{\mathsf m}_{Z{k}}^2)\,}
   \left[\prod_{{w} =1}^{N} 
     \dfrac{{\mathsf m}_{W{w}}^2}
           {\,{\mathsf m}_{W{w}}^2 \!-\! {\mathsf m}_{Z{k}}^2\,}\right]
  \left[ \prod_{{z} \neq {k}}^{K} 
     \dfrac{{\mathsf m}_{Z{z}}^2}
           {\,{\mathsf m}_{Z{z}}^2 \!-\! {\mathsf m}_{Z{k}}^2\,}\right]
\nonumber \\
& \times&
   \left[\prod_{{m} =1}^{M}
     \dfrac{\,{\mathsf m}_{{m}}^2 - {\mathsf m}_{Z{k}}^2\,}{{\mathsf m}_{{m}}^2}\right]
   \left[\prod_{\hat{p} =1}^{p} 
     \dfrac{\,({\mathsf m}_{\hat{p}}^2 - {\mathsf m}_{Z{k}}^2)^2\,}
           {{\mathsf m}_{\hat{p}}^4} \right]\,.
\label{eq:Zh_residue_WW}
\end{eqnarray}

\subsection{General Linear Moose}

Correlation function $[G_{NC}(Q^2)]_{WY}$ as written in eqn. (\ref{gncwy}) has the following heavy pole residue for the general linear moose
\begin{equation}
  [\xi_{Z{k}}]_{WY} =
    e^2 \dfrac{M_Z^2}{{\mathsf m}_{Z{k}}^2- M_Z^2}
    \left[\prod_{{z}\ne {k}}
         \dfrac{{\mathsf m}_{Z{z}}^2}
               {{\mathsf m}_{Z{z}}^2  - {\mathsf m}_{Z{k}}^2 }
    \right]
    \left[\prod_{\hat{p}=1}^{p} 
         \dfrac{{\mathsf m}_{\hat{p}}^2 - {\mathsf m}_{Z{k}}^2}
               {{\mathsf m}_{\hat{p}}^2}
    \right]
    \left[\prod_{\hat{q}=q+1}^{K+1} 
         \dfrac{{\mathsf m}_{\hat{q}}^2 - {\mathsf m}_{Z{k}}^2}
               {{\mathsf m}_{\hat{q}}^2}
    \right].
\end{equation}

\noindent The correlation functions $[G_{NC}(Q^2)]_{WW}$ and $[G_{CC}(Q^2)]_{WW}$ are the same in the general moose as they were in Case I.  Therefore, their heavy pole residues are as given in the previous subsection.

\noindent Applying the consistency relations (\ref{consistency}) to the WY and WW neutral-current heavy pole residues allows us to deduce the YY heavy pole residues:
\begin{eqnarray}
 {[\xi_{Z{k}}^{~}]}_{YY}^{~}  &=& -\dfrac{e^2 (M_W^2 - {\mathsf m}_{Z{k}}^2) M_Z^2}{\,M_W^2(M_Z^2 - {\mathsf m}_{Z{k}}^2 )\,}
  \left[\prod_{{w} =1}^{N} 
     \dfrac{\,{\mathsf m}_{W{w}}^2 \!-\! {\mathsf m}_{Z{k}}^2\,}{{\mathsf m}_{W{w}}^2}\right]
       \left[ \prod_{{z} \neq {k}}^{K} 
     \dfrac{{\mathsf m}_{Z{z}}^2}
           {\,{\mathsf m}_{Z{z}}^2 \!-\! {\mathsf m}_{Z{k}}^2\,}\right]
            \nonumber \\
            &\times &   \left[\prod_{{m} =1}^{M}
     \dfrac{{\mathsf m}_{{m}}^2}{\,{\mathsf m}_{{m}}^2 - {\mathsf m}_{Z{k}}^2\,}\right]
      \left[\prod_{\hat{q}=q+1}^{K+1} 
         \dfrac{({\mathsf m}_{\hat{q}}^2 - {\mathsf m}_{Z{k}}^2)^2}
               {{\mathsf m}_{\hat{q}}^4}
    \right].
    \label{eq:gen-Zh-YY}
\end{eqnarray}

\noindent Alternatively, we can read off the YY heavy pole residues from the correlation function $[G_{NC}(Q^2)]_{YY}$ as written in equation (\ref{eq:exact_YY_related})
\begin{equation}
  [\xi_{Z{k}}]_{YY} =
   - e^2 \dfrac{[{\mathsf m}_{Z{k}}^2 - M_{W'}^2]M_Z^2}
             {M_{W'}^2 [{\mathsf m}_{Z{k}}^2 - M_Z^2]}
      \left[\prod_{{w}=1}^{q-1}
      \dfrac{{\mathsf m}_{W'{w}}^2-{\mathsf m}_{Z{k}}^2}{{\mathsf m}_{W'{w}}^2}
      \right]
      \left[\prod_{{z} \ne {k}}
      \dfrac{{\mathsf m}_{Z{z}}^2}{{\mathsf m}_{Z{z}}^2-{\mathsf m}_{Z{k}}^2}
      \right]
      \left[\prod_{{\hat{q}}=q+1}^{K+1}
      \dfrac{{\mathsf m}_{{\hat{q}}}^2-{\mathsf m}_{Z{k}}^2}{{\mathsf m}_{{\hat{q}}}^2}
      \right]. \nonumber \\
\label{eq:Z_ell_residue_YY_related}
\end{equation}
%


\section{Appendix: Generalized Consistency Conditions}

Using the notation of eqns. (\ref{eq:notationi}) and (\ref{eq:notationii}), the
methods used to derive the consistency relations of sections 6.4 and 8.4 can
be used to prove the following general relations:
\begin{eqnarray}
G^{p,p}_{[0,r]}(Q^2)-G^{p,p}_{[0,q)}(Q^2) & = & 
{\left( G^{p,q}_{[0,r]}(Q^2)\right)^2 \over G^{q,q}_{[0,r]}(Q^2)}~,\qquad
{\rm for}\ 0\le p < q \le r \\
G^{p,p}_{[0,r)}(Q^2)-G^{p,p}_{[0,q)}(Q^2) & = & 
{\left( G^{p,q}_{[0,r)}(Q^2)\right)^2 \over G^{q,q}_{[0,r)}(Q^2)}~,\qquad
{\rm for}\ 0\le p < q  < r \\
G^{p,p}_{(0,r]}(Q^2)-G^{p,p}_{(0,q)}(Q^2) & = & 
{\left( G^{p,q}_{(0,r]}(Q^2)\right)^2 \over G^{q,q}_{(0,r]}(Q^2)}~,\qquad
{\rm for}\ 0< p < q  \le r \\
G^{p,p}_{(0,r)}(Q^2)-G^{p,p}_{(0,q)}(Q^2) & = & 
{\left( G^{p,q}_{(0,r)}(Q^2)\right)^2 \over G^{q,q}_{(0,r)}(Q^2)}~,\qquad
{\rm for}\ 0 < p < q  < r ~.
\end{eqnarray}
%

\section{Appendix: Goldstone Bosons, Duality, and Oscillator Models}

\subsection{Goldstone Bosons}

Consider an arbitrary  $(K+2)$-site linear moose model at $O(p^2)$, given by
\begin{equation}
  {\cal L}_2 =
  \frac{1}{4} \sum_{k=1}^{K+1} f_k^2 \mbox{tr}\left(
    (D_\mu U_k)^\dagger (D^\mu U_k) \right)
  - \sum_{k=0}^{K+1} \dfrac{1}{2} \mbox{tr}\left(
    F^k_{\mu\nu} F^{k\mu\nu}
    \right),
\label{mooselagrangian}
\end{equation}
with
\begin{equation}
  D_\mu U_k = \partial_\mu U_k - i g_{k-1} A^{k-1}_\mu U_k 
                               + i g_k U_k A^{k}_\mu,
\end{equation}
where all  gauge fields $A^k_\mu$ $(k=0,1,2,\cdots, K+1)$ are dynamical and
we have used canonical normalization for the gauge fields. 

The chiral
fields $U_k$ ($k=1,2,\cdots,K+1$) may be written
\begin{equation}
U_k=\exp\left({2i \tilde{\pi}_k \over f_k}\right),\qquad  \tilde{\pi}_k \equiv {\pi^a_k \tau^a \over 2}~,
\end{equation}
where the $\pi^a_k$ are the Goldstone boson fields and the $\tau^a$ are the
usual Pauli matrices. Expanding the sigma-model kinetic terms
eqn. (\ref{mooselagrangian}) to quadratic order in the fields, we find
\begin{equation}
{\cal L}_{KE} = {1\over 2} \sum_{k=1}^{K+1}
\left[\partial_\mu \pi^a_k -{f_k\over 2}(g_{k-1}A^{a,k-1}_\mu - 
g_{k}A^{a,k}_\mu)\right]^2~.
\label{pionke}
\end{equation}
Defining the vectors
\begin{equation}
\vec{\pi}^a = \left(
\begin{array}{c}
\pi^a_1 \\
\pi^a_2 \\
\vdots \\
\pi^a_{K+1}
\end{array}
\right)~, \qquad
\vec{A}^a_\mu = 
\left(
\begin{array}{c}
A^{a,0}_\mu \\
A^{a,1}_\mu \\
\vdots \\
A^{a,K+1}_\mu
\end{array}
\right)~,
\end{equation}
in ``link'' and ``site'' space
we see that the terms in eqn. (\ref{pionke}) may be written
\begin{equation}
{\cal L}_{KE} = {1\over 2}\left[\partial_\mu \vec{\pi}^a - (Q\cdot \vec{A}_\mu)^a\right]^T \cdot
\left[\partial^\mu \vec{\pi}^a - (Q\cdot \vec{A}^\mu)^a\right]~.
\label{pionkee}
\end{equation}
Here the matrix $Q$ is $(K+1) \times (K+2)$ dimensional, and may be written
\begin{equation}
Q=F \cdot D \cdot G~,
\label{decomposition}
\end{equation}
where $F$ is the $(K+1) \times (K+1)$ matrix with diagonal elements $(f_1/2, f_2/2, \ldots, f_{K+1}/2)$, $G$ is the $(K+2) \times (K+2)$ dimensional coupling-constant matrix with diagonal elements
$(g_0, g_1, \ldots, g_{K+1})$, and $D$ is the $(K+1) \times (K+2)$ dimensional difference matrix
\begin{equation}
D = 
\left(
\begin{array}{c|c|c|c|c|c}
1& -1& & & & \\ \hline
 & 1 & -1 & & & \\ \hline
  & & \ddots & \ddots & & \\ \hline
   & & & & 1 & -1 \\
\end{array}
\right)~.
\label{dmatrix}
\end{equation}   
Examining eqn. (\ref{decomposition}), we see that the $(K+2) \times (K+2)$ vector meson mass matrix 
\begin{equation}
{\tiny
M_K^2 = {1\over 4}
\left(
\begin{array}{c|c|c|c|c|c}
g^2_0 f^2_1& -g_0 g_1 f^2_1 & & &  \\ \hline
-g_0 g_1 f^2_1  & g^2_1(f^2_1+f^2_2) & -g_1 g_2 f^2_2 & &   \\ \hline
 & -g_1 g_2 f_2^2 & g^2_2(f^2_2+f^2_3) & -g_2 g_3 f^2_3 &  \\ \hline
 & & \ddots & \ddots & \ddots &   \\ \hline
  & & & -g^{}_{K-1} g^{}_{K} f^2_{K} & g^2_{K}(f^2_{K}+f^2_{K+1}) & -g^{}_{K} g^{}_{K+1} f^2_{K+1} \\ \hline
 & & & & -g^{}_{K} g^{}_{K+1} f^2_{K+1} & g^2_{K+1}f^2_{K+1}\\
\end{array}
\right)~,
}
\label{neutralmatrix}
\end{equation}
may be written
\begin{equation}
M^2_K = Q^T Q~.
\end{equation}

The pion kinetic energy terms in eqn. (\ref{pionkee}) contain those which mix
the Goldstone boson modes with the gauge bosons
\begin{equation}
{\cal L}_{mixing} = -\partial^\mu \vec{\pi}^{aT} \cdot (Q\cdot \vec{A}^a_\mu)~.
\label{mixing}
\end{equation}
Each massive vector boson eigenstate ``eats'' a linear combination of
Goldstone bosons, and the mixing terms are eliminated by the introduction of
gauge-fixing 
\begin{equation}
{\cal L}_\xi = -\,{1\over 2\xi } 
\left[ \partial^\mu \vec{A}^a_\mu + \xi (Q^T\cdot \vec{\pi})^a\right]^T 
\cdot
\left[ \partial^\mu \vec{A}^a_\mu + \xi (Q^T\cdot \vec{\pi})^a\right]~,
\label{gaugefixing}
\end{equation}
shown here for an arbitrary $R_\xi$ gauge. From ${\cal L}_\xi$, we read off the
mass matrix for the unphysical ``eaten'' Goldstone bosons
\begin{equation}
M^2_\xi = \xi Q Q^T \equiv \xi N^2_K~.
\label{pgbmass}
\end{equation}

\subsection{Duality}

Consider the $(K+1) \times (K+1)$ dimensional matrix $N_K^2= Q Q^T$. Inspection of equation
(\ref{decomposition}) shows that $N^2_K$ may be interpreted as the boson mass 
matrix of a ``dual'' moose,
in which the couplings and $F$-constants are interchanged (of course, some overall dimensionful constant will have to be factored out to interpret the $F$'s as ``coupling constants'' -- however, this factor will just be reabsorbed when we form $N$).  Sfetsos \cite{Sfetsos:2001qb}
has shown that every non-zero eigenvalue of $N_K^2$ is also an eigenvalue of $M^2_K$. 

Consider an eigenvector of $N^2_K$, $|\hat{l}_N\rangle$, with eigenvalue $M^2_{\hat{l}} \neq 0$. Note that
$|\hat{l}_N\rangle$ is a $K+1$-dimensional vector. Form the $K+2$ dimensional vector
\begin{equation}
|\hat{l}\rangle = {1\over M_{\hat{l}}} Q^T |\hat{l}_N\rangle~.
\end{equation}
Acting on $|\hat{l}\rangle$ with $M^2_K$ and regrouping terms we see that $|\hat{l}\rangle$ is
an eigenvector of $M^2_K$ with eigenvalue $M^2_{\hat{l}}$.  Correspondingly, given an eigenvector
$|\hat{l}\rangle$ of $M^2_K$ with eigenvalue $M^2_{\hat{l}}\neq 0$, we find that
\begin{equation}
|\hat{l}_N\rangle = {1\over M_{\hat{l}}} Q |\hat{l}\rangle
\label{relation}
\end{equation}
is an eigenvector of $N_K^2$ with the same eigenvalue. Therefore, all nonzero eigenvalues of
$N_K^2$ and $M^2_K$ are equal. 

From eqn. (\ref{pgbmass}) we see that the equality of the non-zero eigenvalues of
$N^2_K$ and $M^2_K$ is equivalent to 
the statement that in 't-Hooft-Feynman gauge, where $\xi = 1$, the masses of the unphysical
Goldstone Bosons are equal to those of the massive vector bosons. Furthermore, comparing
eqns. (\ref{mixing}) and (\ref{relation}), we see that the linear combination of Goldstone Bosons
``eaten'' by a given vector-boson eigenstate is given by the eigenvector of $N^2_K$ with
the same eigenvalue -- and this relationship is true independent of the gauge-fixing
parameter $\xi$.

The results above apply directly to the neutral-boson mass matrix in the general
linear moose model. We note, however,  
that the construction of the unphysical Goldstone boson mass matrix
proceeds analogously  for an arbitrary linear moose, whether or not there are gauge fields at
both endpoints. All that changes is the form of the matrix $D$ in eqn. (\ref{dmatrix}).  
One finds that if the endpoint of the model is gauged the corresponding
endpoint in the dual model corresponds to an global (ungauged) symmetry group,  
and vice versa.  Again, one finds that non-zero eigenvalues for the linear moose 
and its dual are equal. Duality, therefore, applies to the charged-boson mass matrix
as well.

\subsection{Oscillator Models}

Consider a system of $K+2$ masses $m_i$, with adjacent masses coupled by $K+1$ massless
springs with spring constant $k_j$ (the $j$th spring couples masses $m_{j-1}$ to $m_j$). We now
demonstrate that, as
shown by Georgi \cite{Georgi:2004iy}, the eigenfrequencies of this oscillator system are
proportional to the eigenvalues of the vector-boson mass-squared matrix $M^2_K$ if
we choose the masses $m_i \propto 1/g^2_i$ and the spring constants $k_i \propto f^2_i$.

Consider the equation of motion for an arbitrary mass $m_i$:
\begin{equation}
m_i {d^2 x_i \over dt^2} = -k_i(x_i-x_{i-1}) + k_i (x_{i+1}-x_i)~.
\end{equation}
If we introduce the vector
\begin{equation}
\vec{x} = \left(
\begin{array}{c}
x_0 \\
x_1 \\
\vdots \\
x_{K+1}
\end{array}
\right)~,
\end{equation}
the equations for the eigenmodes may be written
\begin{equation}
(D^T \cdot K \cdot D) \cdot \vec{x} = \omega^2\, M\cdot \vec{x}~.
\label{eigenvalue}
\end{equation}
Here $M$ is a $(K+2) \times (K+2)$ dimensional matrix with diagonal
entries $(m_0, m_1, \cdots, m_{K+1})$, $K$ is a $(K+1) \times (K+1)$ dimensional
matrix with diagonal entries $(k_1, k_2, \cdots, k_{K+1})$, and $D$ is the matrix of
eqn. (\ref{dmatrix}). 

Defining $\vec{v} = M^{1/2}\cdot \vec{x}$, we see that the eigenvalue
equation can be written
\begin{equation}
\left[ \tilde{Q}^T \tilde{Q} \right] \vec{v} = \omega^2 \vec{v}~,
\end{equation}
where
\begin{equation}
\tilde{Q} = K^{1\over 2} \cdot D \cdot M^{-{1 \over 2}}~.
\label{ddecomposition}
\end{equation}
Comparing eqns. (\ref{decomposition}) and (\ref{ddecomposition}),
we see that the eigenvalues $\omega^2$ of the spring system correspond to
the vector boson mass-squared values if we make the associations \cite{Georgi:2004iy}
\begin{equation}
K^{1\over 2} \leftrightarrow F~, \qquad M^{-{1\over 2}} \leftrightarrow G~.
\end{equation}

A fixed boundary condition at either end of
the spring system corresponds to taking the mass at that site to infinity, and
hence the corresponding coupling in the linear moose to zero -- corresponding
to a global (ungauged) symmetry group at that site. The duality observed above
for linear moose models implies a relation between oscillator models in which
one exchanges the spring constants with the inverse of the masses.
By reducing the analysis of the mass-squared matrices of the vector bosons
to a coupled spring system, we may use physical intuition to understand which
sets of couplings and $f$-constants that are phenomenologically acceptable.

Finally, we note that the eigenvalue equation (eqn. \ref{eigenvalue}) may
be rewritten
\begin{equation}
(D\cdot M^{-1} \cdot D^T) \cdot (K\cdot D\cdot \vec{x}) = 
\omega^2\, K^{-1}\cdot(K\cdot D \cdot \vec{x})~,
\end{equation}
and we see that all nonzero eigenvalues of the original oscillator
equation are the same in an oscillator where one exchanges
\begin{eqnarray}
M & \leftrightarrow  & K^{-1} \nonumber \\
D & \leftrightarrow & D^T \nonumber \\
\vec{x} & \leftrightarrow & K \cdot D \cdot \vec{x}~.
\label{oscdual}
\end{eqnarray}
These expressions are the oscillator analog of the duality noted in the previous
subsection, and eqn. (\ref{oscdual}) implies that the displacements of the dual oscillator
are related  to the Hooke's law forces experienced by each spring of the original oscillator.


\section{Appendix: Case I with Large ``Bulk'' Coupling}

The analyses presented in this paper are fairly abstract, and it is interesting to consider a special
case in which the results can be checked analytically. In this appendix, we consider
the case of small ``bulk'' coupling, {\it i.e.} $g_{0,K+1} \ll g_m$ for $0<m<K+1$. In the
case $M=0$, and for fermions coupling to the weakly coupled groups at the
ends of the moose ($p=0$, $q=N+1$), this case has been analyzed previously
in ref. \cite{Georgi:2004iy}. In this appendix we will find explicit formulae 
for the masses and ``wavefunctions'' of the $W$ and $Z$ bosons, and directly
compute the parameters $\alpha S$, $\alpha T$, and $\alpha \delta$ to leading nontrivial
order and obtain a connection between the formulae we
have derived previously.

We should note at the outset that although this analysis is interesting for
illustrative purposes, the requirement that the overall coupling be of electroweak strength
(and hence $g_{0,K+1} = {\cal O}(1)$) and that the bulk couplings cannot be too large
($g_m < {\cal O}(4\pi)$) imply that corrections of subleading order are expected to
be of order $10^{-2}$ and are therefore phenomenologically relevant.

\EPSFIGURE[t]{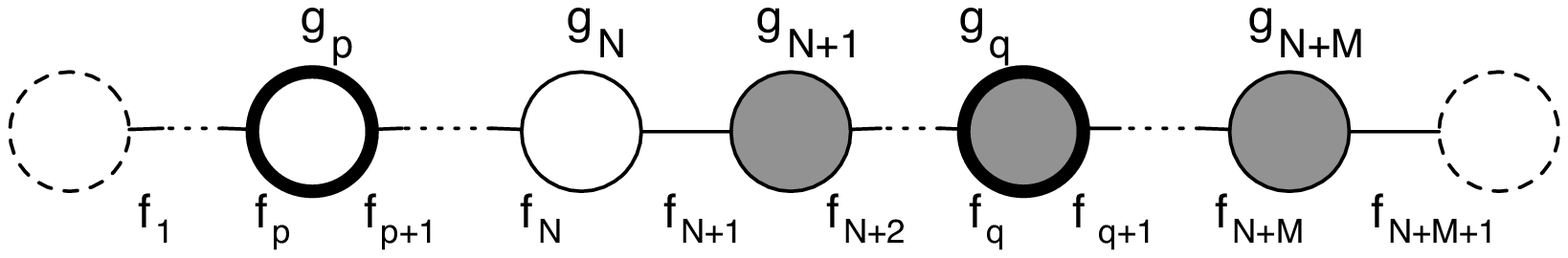,width=0.9\textwidth}
{The moose of fig. \protect{\ref{fig:TheMoose}} in limit $g_{0,K+1} \equiv 0$. The dashed
circles represent the global symmetry groups remaining when $g_{0,K+1} \to 0$.
\label{fig:tzeroglobal}}

\subsection{Approximate Mass Eigenstates}

We begin by considering what happens when $g_{0,K+1} \equiv 0$,
as shown in fig. \ref{fig:tzeroglobal}. The remaining moose has global
symmetry groups on both sides (formerly sites $0$ and $K+1$), and the gauge groups
from sites $1$ to $K$. This moose has a massless particle -- an exact
Goldstone boson with decay constant
\begin{equation}
{1\over F^2_Z} = \sum_{l=1}^{K+1} {1\over f^2_l}~.
\end{equation}
Therefore if we consider sites 0 and K+1 to be weakly gauged, the usual Higgs mechanism causes the mass of the $Z$ boson to be  
\begin{equation}
M^2_Z \approx {g^2_0 + g^2_{K+1} \over 4}\, F^2_Z~,
\end{equation}
to leading order in $g_{0,K+1}$.

If we similarly consider the linear moose for the charged-currents (this includes all of the SU(2) gauged groups, and has the $N+1$ group as global), the charged-current moose has a massless particle -- an exact Goldstone boson with decay constant
\begin{equation}
{1\over F^2_W} = \sum_{l=1}^{N+1}{1\over f^2_l}~,
\end{equation}
and therefore
\begin{equation}
M^2_W \approx {g^2_0 \over 4}\, F^2_W~.
\end{equation}

As in the previous Appendix B, let us define
\begin{equation}
{1\over F^2_i} = \sum_{l=i+1}^{K+1} {1\over f^2_l}\ \ \ ,\ \ 
{1\over \tilde{F}^2_i} = \sum_{l=1}^i {1\over f^2_l}~.
\label{eq:ehs-aa}
\end{equation}
Note that these parameters satisfy
\begin{equation}
{1\over F^2_i} + {1\over \tilde{F}^2_i} \equiv {1\over F^2_Z}\ \ \ ,\ \ 
{1\over F^2_0} = {1\over \tilde{F}^2_{K+1}} \equiv {1\over F^2_Z}~,
\end{equation}
and
\begin{equation}
{1\over \tilde{F}^2_{N+1}} \equiv {1\over F^2_W}~.
\end{equation}

Following the previous calculation, we begin by defining two vectors, 
$|a\rangle$ and $|b \rangle$, in the space of neutral gauge boson
states ($A^{3,0}_\mu, \ldots, A^{3,K+1}_\mu$). We define $|a\rangle$ by
\begin{eqnarray}
\langle i | a \rangle & =  & {g_0 \over g_i}\, {F^2_Z\over F^2_i}, \ \ \ \ i=0, \ldots, K \nonumber \\
\langle i | a \rangle & = & 0~, \ \ \ \ i= K+1
\label{eq:avector}
\end{eqnarray}
and $|b \rangle$ by
\begin{eqnarray}
\langle i | b \rangle & = & 0~, \ \ \ \ i=0 \nonumber \\
\langle i | b \rangle & =  & {g_{K+1} \over g_i}\, {F^2_Z\over \tilde{F}^2_i}, \ \ \ \ i=1, \ldots, K+1~.
\label{eq:bvector}
\end{eqnarray}

The vector $|a\rangle$ has the following properties:  $\langle 0 | a \rangle \equiv 1$,
the amplitudes $\langle i | a\rangle$ for $i=1, \ldots, K$ are suppressed by
$g_0/g_i \ll 1$, and the vector satisfies ``Dirichlet'' boundary conditions ({\it i.e.} the
amplitude vanishes) at site $K+1$. The vector $|b\rangle$ is precisely
the same as $|a\rangle$, exchanging the moose left for right and $g_0$ with $g_{K+1}$. 
Also, to leading order in $g_{0,K+1}/g_i$, these
vectors are normalized and orthogonal to one and other.

Denoting the neutral gauge-boson mass matrix by $M^2_Z$
(unfortunately, the same notation we have used for the $Z$-boson
mass-squared, though the distinction should be clear from
context), we find that
\begin{equation}
\langle a | M^2_Z | a \rangle = {g^2_0 F^2_Z \over 4}
\ \ \ , \ \ \ 
\langle b | M^2_Z | b \rangle = {g^2_{K+1} F^2_Z \over 4}~,
\label{eq:mass1}
\end{equation}
and
\begin{equation}
\langle a| M^2_Z | b\rangle = -\, {g_0 g_{K+1} F^2_Z \over 4}~.
\label{eq:mass2}
\end{equation}

To the order to which we have worked,
\begin{equation}
{1\over e^2} \approx {1\over g^2_0} + {1\over g^2_{K+1}}~,
\end{equation}
so we may define\footnote{The distinction between $\tilde{s}$
defined here, and $s$ as defined in our phenomenological analyses
will become clear in what follows.}
\begin{equation}
g_0 = {e\over \tilde{s}} \ \ \ \ ,\ \ \ \ g_{K+1}={e\over \tilde{c}}~,
\label{eq:couplings}
\end{equation}
where $\tilde{s}^2 + \tilde{c}^2 \equiv 1$.
Using eqns. (\ref{eq:mass1})
and (\ref{eq:mass2}), we immediately see that the
state
\begin{equation}
|\gamma \rangle = \tilde{s}|a\rangle + \tilde{c}|b\rangle~,
\label{eq:gammavector}
\end{equation}
is massless. From eqns. (\ref{eq:current}),
(\ref{eq:avector}), (\ref{eq:bvector}),  (\ref{eq:couplings}), and
(\ref{eq:gammavector}), 
we find that the photon couples to charge, $Q^\mu = J^\mu_3 + J^\mu_Y$,
with strength $e$.

The approximate $Z$ eigenstate is the orthogonal combination
\begin{equation}
|z\rangle = \tilde{c}|a\rangle - \tilde{s}|b\rangle~.
\label{eq:zvector}
\end{equation}
Note that, using eqns. (\ref{eq:mass1}) and (\ref{eq:mass2}), we may explicitly verify
that
\begin{equation}
\langle z | M^2_Z | z \rangle = {g^2_0 + g^2_{K+1} \over 4}\, F^2_Z \equiv 
{e^2 \over 4 \tilde{s}^2 \tilde{c}^2}\, F^2_Z~.
\end{equation}

Similarly, we find that the approximate $W$ eigenstate is given by
\begin{equation}
\langle i | w \rangle  =   {g_0 \over g_i}\, {F^2_W \over F^2_{W,i}}, \ \ \ \ i=0, \ldots, N~,
\end{equation}
where
\begin{equation}
{1\over F^2_{W,i}} = \sum_{l=i+1}^{N+1} {1\over f^2_l}~.
\label{eq:ehs-ab}
\end{equation}
Note that $\langle 0 | w \rangle \equiv 1$, and the amplitude at all other sites
is suppressed by $g_0/g_i \ll 1$.
It is easy to verify that
\begin{equation}
\langle w | M^2_W | w \rangle = {g^2_0 \over 4}\, F^2_W = {e^2\over 4 \tilde{s}^2}\, F^2_W~,
\end{equation}
where $M^2_W$ is the charged-boson mass matrix.

The $Z$ coupling to fermions takes the form
\begin{equation}
{e\over \tilde{s} \tilde{c}}\,  \left( {F^2_Z \over F^2_{W,p}}\, J_{3\mu} - J_{Q\mu} \left[
\tilde{s}^2 - \left(1-{F^2_Z\over F^2_W}\right)\right]\right) Z^\mu ~,
\end{equation}
while the $W$ coupling to fermions is
\begin{equation}
{e\over \sqrt{2} \tilde{s}}{F_W^2 \over {F_{W,p}^2}} J^\mu_\pm W^\mp_\mu~.
\end{equation}
In the above, we have used the relationships 
\begin{eqnarray}
{1\over {\tilde{F}^2_{p}}} &= &{1\over {F_W^2}} - {1\over {F_{W,p}^2}} \\
 {1\over {F_p^2}} &=& {1\over {F_Z^2}} - {1\over {F_W^2}} + {1\over {F_{W,p}^2}}
\end{eqnarray}
which derive from our previous definitions of the $F$-constants in equations (\ref{eq:ehs-aa}) and (\ref{eq:ehs-ab}).

\subsection{Low-Energy Charged- and Neutral-Currents}

Low-energy  $W$-exchange is given by 
\begin{equation}
{\cal L}_{W}  = -\, {2 F_W^2 \over F^4_{W,p}} J^\mu_+ J_{-\mu}
\end{equation}
and exchange of the heavy $W$ bosons contributes approximately
\begin{equation}
{\cal L}_{Wh} = - \frac12 g_p^2 \left\{M_{(0,N+1)}^{-2}\right\}_{p,p} J^\mu_+ J_{-\mu} 
\end{equation}
where we can calculate directly (see Appendix B)
\begin{equation}
\left\{ M_{(0,N+1)}^{-2}\right\}_{p,p} = \frac{4}{g_p^2} \frac{F_W^2}{F_{W,p}^2 \tilde{F}_p^2}.
\end{equation}
Combining these results, we find
\begin{equation}
{\cal L}_{cc} = {\cal L}_W + {\cal L}_{Wh} =  - \frac{2}{F_{W,p}^2} J^\mu_+ J_{-\mu}
\label{eq:ehs-ae}
\end{equation}
so that the definition of $G_F$ is as we expect:
\begin{equation}
\sqrt{2}\,G_F = \frac{1}{F^2_{W,p}}\ .
\end{equation}

The $J^2_3$ portion of low-energy $Z$ exchange is given by
\begin{equation}
{\cal L}_{Z}  = -\, {2 F_Z^2 \over F^4_{W,p}} J^\mu_3 J_{3\mu}
\end{equation}
and the exchange of heavy $Z$ bosons contributes approximately
\begin{eqnarray}
{\cal L}_{Zh} = & - & \frac12 g_p^2 \left\{ M_{(0,K+1)}^{-2}\right\}_{p,p} J^\mu_3 J_{3\mu}   
 -   g_p g_{N+1} \left\{ M_{(0,K+1)}^{-2}\right\}_{p,N+1} J^\mu_3 J_{Y\mu} \nonumber \\
& - &   \frac12 g_{N+1}^2  \left\{ M_{(0,K+1)}^{-2}\right\}_{N+1,N+1}  J^\mu_Y J_{Y\mu} \ .
\label{eq:ehs-ac}
\end{eqnarray}
Computing the matrix elements as in the Appendix B, 
and exchanging $J_Y$ for $J_Q - J_3$, we can rewrite the low-energy effects 
of heavy $Z$ exchange as 
\begin{eqnarray}
{\cal L}_{Zh} = &-& \frac{2}{F_{W,p}^2} \left( 1 - \frac{F_Z^2}{F_{W,p}^2}\right) J^\mu_3 J_{3\mu} 
+ \frac{4}{F_{W,p}^2}  \left(1 - \frac{F^2_Z}{F^2_W}\right) J^\mu_3 J_{Q\mu}  \nonumber \\
&-& \frac{2}{F_{W}^2} \left( 1 - \frac{F_Z^2}{F_{W}^2}\right) J^\mu_Q J_{Q\mu}\ .
\label{eq:ehs-ad}
\end{eqnarray}
Note that if we combine the $J_3^2$ pieces of equations (\ref{eq:ehs-ac}) and (\ref{eq:ehs-ad}) we find the total low-energy exchange in this channel is
\begin{equation}
 - \frac{2}{F_{W,p}^2} J^{3\mu} J_{3\mu}
\end{equation}
and comparing this with equation (\ref{eq:ehs-ae}) shows that $\rho = 1$. 
In the limit $N+1=K+1$ isospin is a good symmetry, $F_W = F_Z$ and the contributions
from heavy boson exchange (${\cal L}_{Wh,Zh}$) are proportional to $\vec{J^\mu}^2$. 

For a phenomenologically viable model, we must require that the contributions from heavy boson exchange are small
\begin{equation}
0<\left({F^2_{W,p}\over F^2_W} - 1\right) \equiv A  \ll 1 \ \ \ \&\ \ \ 
0< \left({F^2_W \over F^2_Z} - 1\right) \equiv B \ll 1~.
\end{equation}
In this case we may expand the low-energy
four-fermion operator in eqn. (\ref{eq:ehs-ad}) to lowest order in $A$ and $B$. Doing so,
we find
\begin{equation}
{\cal L}_{Zh} \approx -\,{2 \over F^2_Z}  \left({F^2_{W,p}\over F^2_W} - 1\right) J^\mu_3 J_{3\mu}
-\,{2\over F^2_Z} \left({F^2_W \over F^2_Z} - 1\right) J^\mu_Y J_{\mu Y}~.
\end{equation}
Comparing this Lagrangian to the forms of the amplitudes, we find
\begin{eqnarray}
\alpha T & = & - \left({F^2_W \over F^2_Z} - 1\right) \label{eq:tfour} \\
\alpha \delta & = &  4 s^2 c^2 \left({F^2_{W,p}\over F^2_W} - 1\right)>0~.
\label{eq:dfour}
\end{eqnarray}
In the limit $p=0$, $F_{W,p} = F_W$ and the contributions from
heavy boson exchange are proportional
to $J^\mu_Y J_{Y\mu}$ as expected in the case of large isospin-violation. 

\subsection{Z-pole Observables}

The general analyses  the formulae for $\alpha S$, $\alpha T$, and $\alpha \delta$ 
are given by  eqns. (\ref{eq:SSi}), (\ref{eq:TTi}), and (\ref{eq:deltai}):
\begin{eqnarray}
\alpha S &=& 4 s_z^2 c_z^2 M_Z^2 \left[ \Sigma_Z - \Sigma_{{p}} - \Sigma_{{q}}\right]~, \\
\alpha T &=& s_z^2 M_Z^2 \left[\Sigma_Z - \Sigma_W + \Sigma_{M} - 2 \Sigma_{{q}}\right]~, \\
\alpha \delta & = & - 4 s^2_Z c^2_Z M^2_W \left[ \Sigma_W - \Sigma_r -\Sigma_p\right]~.
\end{eqnarray}
To leading order in the strong bulk coupling expansion all of the mass-squareds appearing
in $\Sigma_W$, $\Sigma_Z$, and $\Sigma_r$ are large -- and are
proportional to squares of the bulk couplings. On the other hand, the $\Sigma_{p,q}$
have contributions proportional to $1/g^2_{0,K+1}$:
\begin{eqnarray}
\Sigma_{{p}} &=& \frac{4}{g_0^2} \left( \frac{1}{F_W^2} - \frac{1}{F_{W,p}^2}\right) \\
\Sigma_{{M}} &=& \frac{4}{g_{K+1}^2} \left(\frac{1}{F_Z^2} - \frac{1}{F_{W}^2}\right)
\end{eqnarray}
while, since this is a model with $q=N+1$, we also have $\Sigma_{{M}} = \Sigma_{{q}}$.
As a result, we find\footnote{To this order,  we may substitute $s_Z$ and $c_Z$ for
$\tilde{s}$ and $\tilde{c}$, and $F_W$ for $F_Z$, in the 
coefficients of  the expressions for $\alpha T$ and $\alpha \delta$.
Also, for $A,\, B \ll 1$, 
\begin{equation}
\left({F^2_W \over F^2_Z}-1\right) \approx \left(1-{F^2_Z \over F^2_W}\right)~,
\ \ \ \ 
\left({F^2_{W,p}\over F^2_W} -1 \right) \approx \left(1-{F^2_W \over F^2_{W,p}}\right)~,
\end{equation}
and the values of $\alpha T$ 
and $\alpha \delta$ are in agreement with the low-energy analysis of the previous section.}
\begin{eqnarray}
\alpha S &=& - 4 F_Z^2 \left[ \tilde{s}^2 \left( \frac{1}{F_W^2} - \frac{1}{F_{W,p}^2} \right)
+ \tilde{c}^2 \left( \frac{1}{F_Z^2} - \frac{1}{F_W^2} \right) \right]~,\\
\alpha T &=& \left( 1 - \frac{F_W^2}{F_Z^2}\right) < 0 ~,\\
\alpha \delta & = & 4 \tilde{s}^2 \tilde{c}^2 \left(1-{F^2_W \over F^2_{W,p}}\right)~, 
\end{eqnarray}
with the consequences
\begin{equation}
\alpha S - 4 c^2 \alpha T \approx - 4 F_W^2 \tilde{s}^2 
\left( \frac{1}{F_W^2} - \frac{1}{F_{W,p}^2} \right) < 0
\end{equation}
and
\begin{equation}
\hat{S} \propto \alpha S - 4 c^2 \alpha T + {\alpha \delta \over c^2} \simeq 0~.
\end{equation}
As shown by eqn. (\ref{eq:shatgeneral}), the next order corrections to $\hat{S}$ 
are known to be positive. 

Note that the requirement that $\alpha \delta$
be small implies that the constants $f_i$, $0\le i \le p$, must be large.



\end{document}